\documentclass{article}

\usepackage[preprint]{neurips_2025}

\usepackage[utf8]{inputenc} 
\usepackage[T1]{fontenc}    
\usepackage{hyperref}       
\usepackage{url}            
\usepackage{booktabs}       
\usepackage{amsfonts}       
\usepackage{nicefrac}       
\usepackage{microtype}      
\usepackage{xcolor}         
\usepackage{bbding}         
\usepackage{multirow}       
\usepackage{longtable}      
\usepackage{caption}        
\usepackage{graphicx}       
\usepackage{makecell}       
\usepackage{bm}             
\usepackage{CJKutf8}        
\usepackage{threeparttable} 
\usepackage{wrapfig}        
\usepackage{amsmath}        
\usepackage{algorithm}      
\usepackage{algpseudocode}  

\title{
\begin{minipage}{0.1\textwidth}
\includegraphics[width=1.\linewidth]{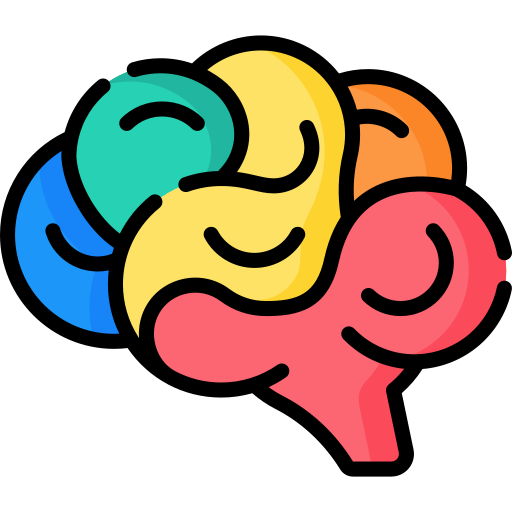}
\end{minipage}
\begin{minipage}{0.8\textwidth}
BrainStratify: Coarse-to-Fine Disentanglement of Intracranial Neural Dynamics
\end{minipage}
}

\author{Hui Zheng\textsuperscript{2,4}, Hai-Teng Wang\textsuperscript{1}, Yi-Tao Jing\textsuperscript{3}, Pei-Yang Lin\textsuperscript{1}, Han-Qing Zhao\textsuperscript{2,4},\\
\textbf{Wei Chen\textsuperscript{1}, Peng-Hu Wei\textsuperscript{5}, Yong-Zhi Shan\textsuperscript{5}, Guo-Guang Zhao\textsuperscript{5}, Yun-Zhe Liu\textsuperscript{\dag,1,4}}\\
\textsuperscript{1}Beijing Normal University, \textsuperscript{2}Peking University, \textsuperscript{3}Institute of Automation, Chinese Academy of Sciences\\
\textsuperscript{4}Chinese Institute for Brain Research, \textsuperscript{5}Capital Medical University, Xuanwu Hospital, Beijing\\
\texttt{\dag yunzhe.liu@bnu.edu.cn}
}

\begin{document}

\maketitle

\begin{abstract}
  Decoding speech directly from neural activity is a central goal in brain-computer interface (BCI) research. In recent years, exciting advances have been made through the growing use of intracranial field potential recordings, such as stereo-ElectroEncephaloGraphy (sEEG) and ElectroCorticoGraphy (ECoG). These neural signals capture rich population-level activity but present key challenges: (\romannumeral1) task-relevant neural signals are sparsely distributed across sEEG electrodes, and (\romannumeral2) they are often entangled with task-irrelevant neural signals in both sEEG and ECoG. To address these challenges, we introduce a unified Coarse-to-Fine neural disentanglement framework, BrainStratify, which includes (\romannumeral1) identifying functional groups through spatial-context-guided temporal-spatial modeling, and (\romannumeral2) disentangling distinct neural dynamics within the target functional group using Decoupled Product Quantization (DPQ). We evaluate BrainStratify on two open-source sEEG datasets and one (epidural) ECoG dataset, spanning tasks like vocal production and speech perception. Extensive experiments show that BrainStratify, as a unified framework for decoding speech from intracranial neural signals, significantly outperforms previous decoding methods. Overall, by combining data-driven stratification with neuroscience-inspired modularity, BrainStratify offers a robust and interpretable solution for speech decoding from intracranial recordings.
\end{abstract}

\section{Introduction}
Intracranial neural signals refer to the biometric information collected from the brain through invasive recording techniques (e.g., stereo-ElectroEncephaloGraphy (sEEG) and ElectroCorticoGraphy (ECoG)). Their patterns provide rich and high-resolution insights toward understanding the physiological functions of the brain and the mechanism of related diseases, leading to various applications including speech decoding \cite{moses2021neuroprosthesis,metzger2023high,zheng2025discrete,chau2024population}, motor intention decoding \cite{natraj2025sampling,silversmith2021plug}, neurological disorders detection \cite{zhang2023brant,yuan2023ppi,li2025deep}, among others. Although many studies \cite{moses2021neuroprosthesis,zheng2025discrete,chau2024population} have recently shown promising results in speech decoding (e.g., vocal production and speech perception) based on sEEG and ECoG, significant challenges in modeling intracranial neural signals remain unresolved.

\begin{wrapfigure}{r}{0.4\linewidth}
  \centering
  \includegraphics[width=\linewidth]{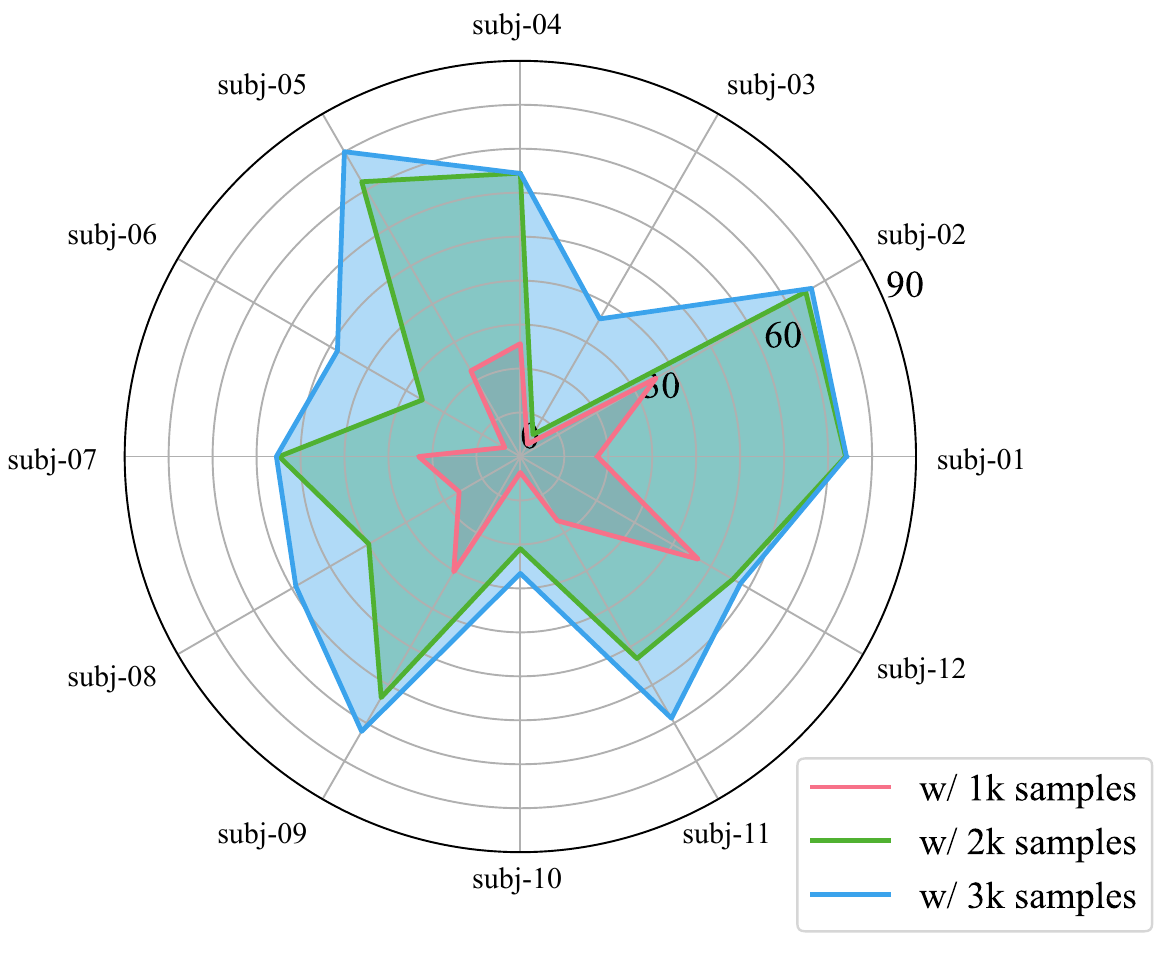}
  \caption{The 61-word performance on Du-IN \cite{zheng2025discrete} dataset using top-10 channels selected via the MC strategy across varying numbers of labeled samples.}
  \label{fig:mcsf}
\end{wrapfigure}

Compared to ECoG, sEEG provides more depth information from specific brain regions, making it particularly attractive in both brain-computer interface (BCI) applications \cite{chau2024population,zheng2025discrete,mentzelopoulos2024neural} and fundamental neuroscience studies \cite{subramaniam2024revealing,norman2019hippocampal,domenech2020neural}. Despite their potential, sEEG recordings present a unique challenge. In practice, sEEG electrodes are sparsely distributed across the brain, requiring researchers to first identify task-relevant channels before decoding \cite{wang2023brainbert,mentzelopoulos2024neural,zheng2025discrete}. For instance, BrainBERT \cite{wang2023brainbert} adopts a single-channel (SC) strategy, independently evaluating and ranking channels based on decoding performance. In contrast, Du-IN \cite{zheng2025discrete} utilizes a multi-channel (MC) approach, analyzing all channels collectively and ranking them based on the learned weight distribution. Both methods require substantial labeled data for supervision, posing significant challenges, as large-scale labeling in medical experiments is often prohibitively costly or unfeasible. As shown in Figure \ref{fig:mcsf}, when labeled data are scarce, the selected channels often fail to align with those containing the target neural activity.

In addition, another major challenge arises from the nature of the intracranial recordings themselves. Invasive recording techniques (e.g., sEEG, ECoG) capture aggregated neural activity from populations of neurons (i.e., population-level recordings \cite{chau2024population}). While sEEG can enhance spatial resolution through techniques like bi-polar (or Laplacian) re-reference \cite{li2018optimal}, intracranial neural signals inherently represent a mixture of signals from multiple neural dynamics. Without explicit mechanisms to disentangle distinct neural dynamics within specific brain regions, models struggle to extract fine-grained states from intracranial neural signals, leading to reduced decoding performance.

To tackle these issues, we propose BrainStratify, a general framework for decoding speech from intracranial neural signals. This framework comprises two complementary stages: (1) Coarse Disentanglement Learning and (2) Fine Disentanglement Learning. In the Coarse-DL stage, we pre-train a temporal-spatial model with a spatial context task and then cluster channels into functional groups \cite{buzsaki2006rhythms,silva2024speech} based on the sparse inter-channel attention graph. In the Fine-DL stage, we utilize Decoupled Product Quantization (DPQ) to disentangle different neural dynamics \cite{chapeton2022micro,mudrik2024sibblings,mudrik2025creimbo} within target functional groups, enhancing the identification of fine-grained task-relevant states.

To validate the effectiveness of our proposed framework, we evaluate BrainStratify on two publicly available sEEG datasets \cite{zheng2025discrete,wang2024brain} and our collected word-reading (epidural) ECoG dataset (see Appendix \ref{sec:supp-expr-design} for more details). Empirically, BrainStratify outperforms existing channel clustering methods \cite{chen2025similarity,qiu2024duet}, identifying channel cluster that faithfully aligns with those containing target neural activity. In addition, BrainStratify achieves SOTA performance in all speech decoding tasks, particularly excelling in word decoding tasks \cite{zheng2025discrete}. This success stems from its ability to effectively integrate distinct modular neural dynamics within the target functional group -- a critical requirement for decoding complex, interdependent neural patterns.

To sum up, the main contributions of our work comprise:
\begin{enumerate}
  \item \textbf{Coarse-to-Fine neural disentanglement:} We propose a unified neural disentanglement framework, BrainStratify, that identifies functional channel groups and disentangles neural dynamics within target functional groups, through two complementary stages.
  \item \textbf{Neuroscience-inspired design:} BrainStratify leverages neuroscience insights (e.g., functional brain networks, modular neural dynamics) in its architecture, discovering task-relevant channel groups based on the sparse inter-channel attention graph learned via self-supervision.
  \item \textbf{State-of-the-art (SOTA) performance:} Our framework achieves SOTA performance in decoding speech from intracranial neural signals (e.g., sEEG, ECoG) across multiple datasets, demonstrating robust effectiveness across diverse neural recording modalities.
\end{enumerate}

\section{Related Works}
\subsection{Self-supervised Learning in BCI}
Recently, the pre-trained temporal-spatial models (i.e., foundation models) have drawn significant attention across diverse neural recording modalities, including EEG \cite{jiang2024large,wang2024eegpt,wang2024cbramod}, sEEG \cite{wang2023brainbert,zhang2023brant,chau2024population}, fMRI \cite{caro2023brainlm,dong2024brain}, and so on. Due to their adaptability in spatial modeling, these models robustly handle varying numbers of input channels and excel at channel-level classification tasks (e.g., epilepsy detection). PopT \cite{chau2024population} further utilizes \texttt{[CLS]} token to aggregate channels, achieving superior performance in decoding modular cognitive states (e.g., sentence onset detection) of speech perception.

Other approaches \cite{zheng2025discrete,wu2024towards} fuse the manually selected channels into region-level tokens and then pre-train temporal models based on them. While these methods perform well in decoding complex cognitive states (e.g., vocal production), their effectiveness depends on whether the manually selected channels faithfully represent the target functional groups.

\subsection{Channel Cluster in Time Series}
Channel clustering methods, which leverage cluster information instead of individual channel identities, have gained significant attention in Multivariate Time Series Forecasting (MTSF) research \cite{chen2025similarity,qiu2024duet,huang2023crossgnn}. CCM \cite{chen2025similarity} employs static cluster embeddings and a cross-attention mechanism to cluster channels into different groups, enhancing time series forecasting tasks. Considering the dynamic nature of time series, DUET \cite{qiu2024duet} utilizes correlation-based metric learning to capture the relationship among channels, which are then integrated via masked attention for downstream forecasting tasks.

All existing channel cluster methods for time series analysis primarily rely on low-level correlations in raw time series, which are unsuitable for intracranial neural signals. While these signals reflect aggregated neural activity from neuronal populations, similar firing patterns hold distinct functional meanings by location \cite{buzsaki2006rhythms}, producing inherently multimodal signals across channels. This suggests that correlation-based clustering may inadequately capture functionally relevant neural clusters.

\subsection{Disentanglement Representation Learning}
Disentangling independent latent components from observations is a desirable goal in representational learning \cite{higgins2017beta,locatello2019challenging,wang2024disentangled}, with numerous applications in fields such as computer vision \cite{hsu2023disentanglement,hsu2024tripod}, time series analysis \cite{oublal2024disentangling,woo2022cost}, and neuroscience \cite{zhou2020learning,wang2024exploring,li2025revisit}. QLAE \cite{hsu2023disentanglement} leverages learnable latent codex \cite{van2017neural} combined with weight decay regularization \cite{loshchilov2017decoupled} to extract human-interpretable representations from raw images. Tripod \cite{hsu2024tripod} further enhances disentanglement along [width, height] dimensions by introducing minimal mixed generator derivatives to guide feature separation.

Unlike image data, which typically requires disentangling features along [width, height] dimensions, neural data often requires disentanglement along the channel dimension -- analogous to how RGB channels are treated in images -- rather than the temporal dimension. pi-VAE \cite{zhou2020learning} incorporates supervision labels to model the relation between the latent and task variables simultaneously. PDisVAE \cite{li2025revisit} encourages group-wise independence in learned representations via partial-correlation constraint, tackling scenarios where multiple factors are non-separably entangled.

\section{Method}
\begin{figure}[h]
  \centering
  \includegraphics[width=\linewidth]{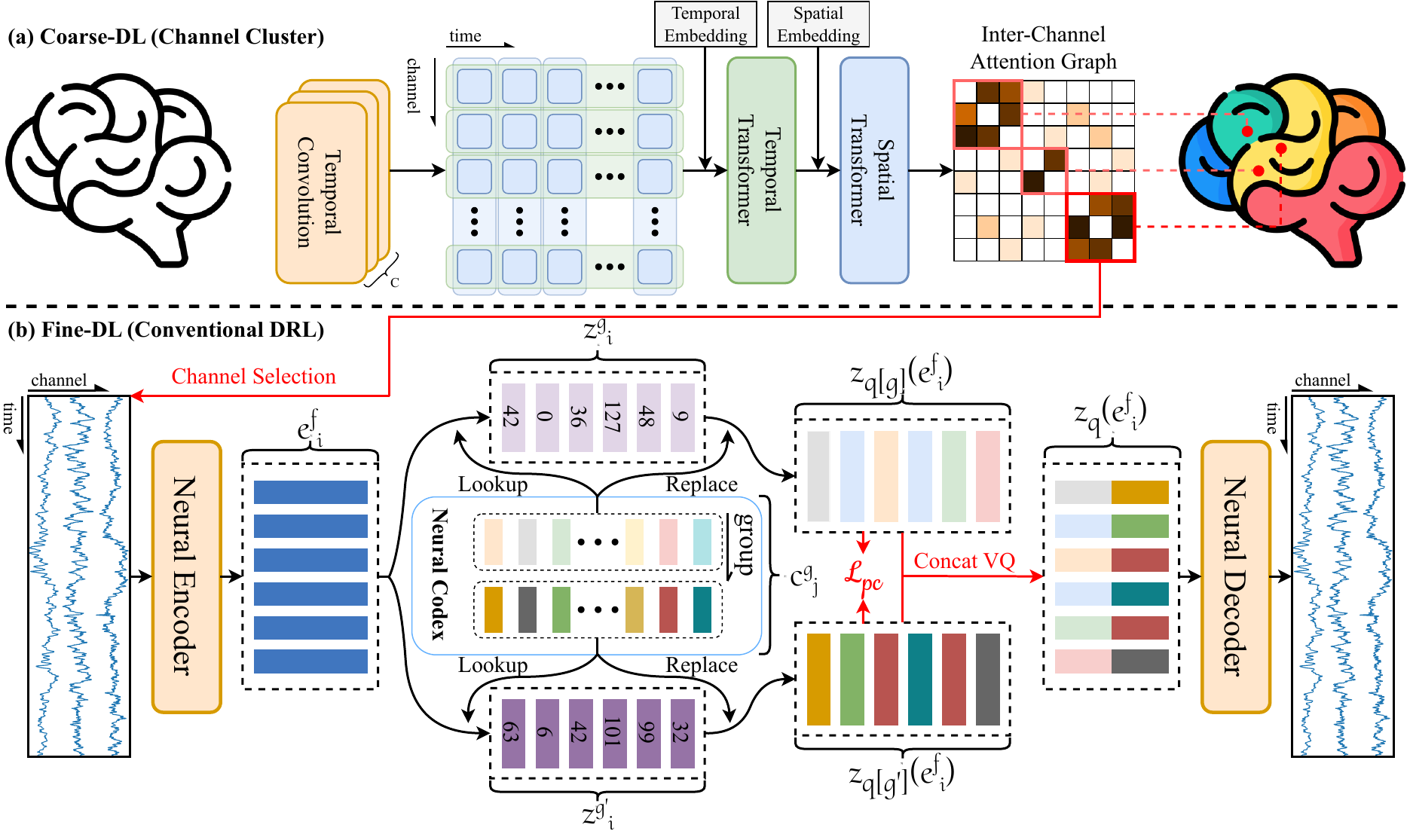}
  \caption{\textbf{Overview of BrainStratify framework.} \textbf{(a).} Coarse Disentanglement Learning Stage (BrainStratify-Coarse). \textbf{(b).} Fine Disentanglement Learning Stage (BrainStratify-Fine).}
  \label{fig:cfd}
\end{figure}

The overall architecture of BrainStratify is illustrated in Figure \ref{fig:cfd}, where the framework contains two complementary stages: (1) Coarse Disentanglement Learning Stage and (2) Fine Disentanglement Learning Stage (i.e., Disentanglement Representation Learning (DRL)).

\subsection{Task Definition}
The evaluation spans three intracranial neural datasets, including both sEEG and ECoG neural signals. In the Du-IN dataset \cite{zheng2025discrete}, we assess both the original 61-word classification (CLS) task and the 49-syllable connectionist temporal classification (CTC) task, which follows Willett et al. \cite{willett2023high,fan2023plug}. In addition, we follow the experimental design outlined by Du-IN \cite{zheng2025discrete} to collect an (epidural) ECoG dataset. In our collected dataset, we evaluate the 62-word CLS task and the 49-syllable CTC task. In the Brain Treebank dataset \cite{wang2024brain}, we evaluate four 2-state CLS tasks (e.g., sentence onset detection).

Multi-channel sEEG signals are represented as $\mathcal{X}\in\mathbb{R}^{C\times T}$, where $C$ is the number of channels and $T$ is the total timestamps. For CLS tasks, the paired label is $\bm{y}\in \mathcal{Y}$, where $\mathcal{Y}$ represents the word-set (or state-set). For CTC tasks, the paired label sequence is $\bm{y}=\{\bm{y}_{i}\in \mathcal{Y}|i=1,...,L\}$, where $\mathcal{Y}$ represents the syllable-set and $L$ is the length of syllable sequence. See Appendix \ref{sec:supp-model-details} for more details.

\subsection{Coarse Disentanglement Learning}
We introduce BrainStratify-Coarse (Figure \ref{fig:cfd} (a)), a general architecture for sEEG-based functional group identification. BrainStratify-Coarse contains three parts: (1) Patch Tokenizer, (2) Temporal \& Spatial Transformer, and (3) Channel Cluster Module. Following brain signal foundation models \cite{zhang2023brant,jiang2024large,wang2024cbramod}, we segment the sEEG signals into patches, enabling flexible input processing. For each sample $\mathcal{X}$, we use a $W_{c}$-length window (0.25s) without overlap, obtaining $\mathcal{X}^{c}_{p}=\{\bm{x}^{c}_{i,j}\in\mathbb{R}^{W_{c}}|i=1,...,C;j=1,...,N_{c}\}$, where $N_{c}=\lfloor\frac{T}{W_{c}}\rfloor$ and the number of patches is $|\mathcal{X}^{c}_{p}|=C\times N_{c}$.

\paragraph{Patch Tokenizer.}Since sEEG signals from different channels are inherently partial observations of whole-brain activity, channel-specific patch tokenizers are employed to map signals into a unified latent brain space. Each tokenizer encodes patches with stacked convolution blocks. Each block contains a 1D convolution, group normalization \cite{wu2018group}, and GELU activation \cite{hendrycks2016gaussian}. The patch embeddings are $\mathcal{E}^{c}_{p}=\{\bm{e}^{c}_{i,j}\in\mathbb{R}^{d}|i=1,...,C;j=1,...,N_{c}\}$, where $d$ is the dimension of embeddings.

\paragraph{Temporal \& Spatial Transformer.}We add the parameter-free temporal embeddings introduced in \cite{vaswani2017attention} to inject the temporal information. Then, the embeddings are fed into "Temporal Transformer" \cite{vaswani2017attention} to get the temporal-transformed embeddings $\mathcal{E}^{c}_{t}$. After that, we add either learnable or MNI-based \cite{chau2024population} spatial embeddings to inject the spatial information. Finally, the embeddings are fed into "Spatial Transformer" to get the spatial-transformed embeddings $\mathcal{E}^{c}_{s}$.

\paragraph{Spatial Context Pre-training.}To avoid shortcuts in mask-based reconstruction tasks \cite{zhang2023brant,wang2024cbramod}, where models might over-rely on intra-channel temporal patterns, we adapt the spatial context task from PopT \cite{chau2024population}, refining it for efficient convergence with limited data. Given a 4-second sEEG sample $\mathcal{X}\in\mathbb{R}^{C\times T}$, 10 \% of channels are randomly selected to have their activity replaced by activity from unrelated time points. The model is trained to detect discrepancies between the spatial-transformed embeddings $\mathcal{E}^{c}_{s}$ and the temporal-transformed embeddings $\mathcal{E}^{c}_{t}$ before spatial attention:
\begin{equation}
  \label{equ:spatial-context-loss}
  \mathcal{L}_{c}=\mathrm{BCE}(\mathrm{Linear}(||\mathcal{E}^{c}_{s}-\mathcal{E}^{c}_{t}||^{2}_{2})).
\end{equation}
With this adaptation, the model converges rapidly to $\geq$ 95\%, even with only 1-hour sEEG data.

\paragraph{Channel Cluster Module.}We forward the pre-trained model with all sEEG samples and compute channel connectivity $\mathcal{P}\in\mathbb{R}^{C\times C}$ by aggregating attention matrices from "Spatial Transformer" across layers, temporal patches, and samples. Spectral cluster \cite{ng2001spectral} is applied to group channels into functional clusters, as shown in Figure \ref{fig:attn-graph} (c). Other cluster approaches (e.g., hierarchical cluster) yield similar results. The channels within each functional group typically provide complementary information, collectively representing a complete and complex neural function. We select and combine these groups based on their performance in specific downstream tasks.

\subsection{Fine Disentanglement Learning}
We present BrainStratify-Fine (Figure \ref{fig:cfd} (b)), a general framework for decoding speech from intracranial neural signals. Like Du-IN \cite{zheng2025discrete}, BrainStratify-Fine employs a two-stage pre-training pipeline (i.e., VQ-VAE and MAE stage). To identify fine-grained states from intracranial neural signals, we propose Decoupled Product Quantization (DPQ), which disentangles distinct neural dynamics within target functional groups. DPQ integrates product quantization \cite{jegou2010product} with partial-correlation constraint \cite{hazarika2020misa,li2025revisit}, enforcing group-wise independence along the channel dimension.

BrainStratify-Fine adopts the architecture of Neural Encoder in Du-IN \cite{zheng2025discrete}, which contains two parts: (1) Patch Tokenizer and (2) Temporal Transformer. Unlike the patch process during Coarse Disentanglement Learning, we segment neural signals into temporal patches along the temporal dimension. For each sample $\mathcal{X}$, we use a $W_{f}$-length window (0.15s) with $S_{f}$-length stride (0.1s), obtaining $\mathcal{X}^{f}_{p}=\{\bm{x}^{f}_{i}\in\mathbb{R}^{C\times W_{f}}|i=1,...,N_{f}\}$, where $N_{f}=\lfloor\frac{T}{S_{f}}\rfloor$ is the number of patches.

\paragraph{Patch Tokenizer.}The patch tokenizer comprises a linear projection and stacked convolution blocks. Each block contains a 1D convolution, group normalization \cite{wu2018group}, and GELU activation \cite{hendrycks2016gaussian}. The patch embeddings are $\mathcal{E}^{f}_{p}=\{\bm{e}^{f}_{i}\in\mathbb{R}^{d}|i=1,...,N_{f}\}$, where $d$ is the dimension of embeddings.

\paragraph{Temporal Transformer.}We add temporal embeddings \cite{vaswani2017attention} to inject temporal information. Then, the embeddings are fed into "Temporal Transformer" to get the temporal-transformed embeddings $\mathcal{E}^{f}$. For downstream evaluations, we add a classification head to support CLS and CTC tasks.

\paragraph{Decoupled Product Quantization.}Before pre-training BrainStratify-Fine through mask modeling (i.e., MAE stage), we discretize neural tokens into discrete codes during the VQ-VAE stage. To identify fine-grained functional modules \cite{chapeton2022micro,mudrik2024sibblings,mudrik2025creimbo} within target functional groups, we propose Decoupled Product Quantization (DPQ), using multiple codexes to capture distinct neural dynamics.

The output embeddings $\mathcal{E}^{f}=\{\bm{e}^{f}_{i}\in\mathbb{R}^{d}|i=1,...,N_{f}\}$ from "Neural Encoder" are fed into a vector quantizer, which consists of $G$ parallel sub-quantizers (i.e., neural codexes). The $g$-th neural codex is defined as $\mathcal{C}_{g}=\{\bm{c}^{g}_{j}|j=1,...,N_{codex}\}\in\mathbb{R}^{N_{codex}\times d_{codex}}$, where $N_{codex}$ is the number of discrete codes and $d_{codex}$ is the dimension of code embeddings. We utilize a linear projection $\bm{\mathrm{z}}_{c[g]}$ to get the mapped embeddings $\bm{\mathrm{z}}_{c[g]}(\mathcal{E}^{f})=\{\bm{\mathrm{z}}_{c[g]}(\bm{e}^{f}_{i})\in\mathbb{R}^{d_{codex}}|i=1,...,N_{f}\}$ in the codex space. Then, the codex looks up the nearest neighbor of each embedding $\bm{\mathrm{z}}_{c[g]}(\bm{e}^{f}_{i})$ in the neural codex $\mathcal{C}_{g}$.
\begin{equation}
\begin{split}
  &\bm{\mathrm{z}}_{q[g]}(\mathcal{E}^{f})=\{\bm{\mathrm{z}}_{q[g]}(\bm{e}^{f}_{i})|i=1,...,N_{f}\},\\
  &\bm{\mathrm{z}}_{q[g]}(\bm{e}^{f}_{i})=\bm{c}^{g}_{z^{g}_{i}},\quad z^{g}_{i}=\mathop{\arg\min}\limits_{j}||\ell_{2}(\bm{\mathrm{z}}_{c[g]}(\bm{e}^{f}_{i}))-\ell_{2}(\bm{c}^{g}_{j})||_{2},
\end{split}
\end{equation}
where $\ell_{2}$ represents $\ell_{2}$ normalization and $\bm{\mathrm{z}}_{q[g]}(\bm{e}^{f}_{i})$ is the quantized vector from $g$-th sub-quantizer. As shown in Figure \ref{fig:cfd} (b), $\bm{\mathrm{z}}_{q[g]}(\bm{e}^{f}_{i})$ from $G$ sub-quantizers are concatenated to the full code $\bm{\mathrm{z}}_{q}(\bm{e}^{f}_{i})=\left[\bm{\mathrm{z}}_{q[1]}(\bm{e}^{f}_{i}),...,\bm{\mathrm{z}}_{q[G]}(\bm{e}^{f}_{i})\right]$. Then, the code $\bm{\mathrm{z}}_{q}(\bm{e}^{f}_{i})$ is linearly mapped to the quantized embedding $\bm{z}_{i}\in\mathbb{R}^{d}$, which is equivalent to summing $\bm{z}^{q[g]}_{i}\in\mathbb{R}^{d}$ from $G$ sub-quantizers, i.e., $\bm{z}_{i}=\sum^{G}_{g=1}\bm{z}^{q[g]}_{i}$.

Given the quantized embeddings $\mathcal{Z}=\{\bm{z}_{i}|i=1,...,N_{f}\}$, the Neural Decoder converts them back into neural signals $\tilde{\mathcal{X}}^{f}_{p}=\{\tilde{\bm{x}}^{f}_{i}|i=1,...,N_{f}\}$. The mean squared error (MSE) loss is utilized to guide the regression. Besides, we introduce partial-correlation constraint \cite{hazarika2020misa,li2025revisit} to encourage group-wise independence. The total loss $\mathcal{L}_{f}^{\mathcal{VQ}}$ for training the VQ-VAE model is:
\begin{equation}
\begin{split}
  &\mathcal{L}_{f}^{\mathcal{VQ}}=\sum_{i=1}^{N_{f}}\left[\mathcal{L}_{rgs}+\mathcal{L}_{vq}+\mathcal{L}_{pc}\right],\quad
  \mathcal{L}_{rgs}=||\tilde{\bm{x}}^{f}_{i}-\bm{x}^{f}_{i}||_{2}^{2},\quad
  \mathcal{L}_{pc}=\sum_{j=1}^{G-1}\Big[\sum_{k=j+1}^{G}\bm{z}^{q[j]}_{i}\cdot \bm{z}^{q[k]}_{i}\Big],\\
  &\mathcal{L}_{vq}=\sum_{g=1}^{G}\Big[||\bm{\mathrm{sg}}[\bm{\mathrm{z}}_{c[g]}(\bm{e}^{f}_{i})]-\bm{\mathrm{z}}_{q[g]}(\bm{e}^{f}_{i})||_{2}^{2}+\beta||\bm{\mathrm{z}}_{c[g]}(\bm{e}^{f}_{i})-\bm{\mathrm{sg}}[\bm{\mathrm{z}}_{q[g]}(\bm{e}^{f}_{i})]||_{2}^{2}\Big],
\end{split}
\end{equation}
where $\bm{\mathrm{sg}}$ represents the stop-gradient operation, which is an identity at the forward pass and has zero gradients. To stabilize the codex update, we use the exponential moving average strategy \cite{van2017neural}.

\paragraph{DPQ-guided Mask Modeling.}BrainStratify-Fine uses DPQ-guided mask modeling to learn contextual representations. Given a sample $\mathcal{X}$, the patch tokenizer transforms it into patch embeddings $\mathcal{E}^{f}_{p}$. Around 50\% of embeddings are patch-wise chosen and masked. The masked position is termed as $\mathcal{M}$. Then, a shared learnable embedding $\bm{e}_{[M]}\in\mathbb{R}^{d}$ is used to replace the original patch embeddings:
\begin{equation}
  \mathcal{E}^{f}_{m}=\{\bm{e}^{m}_{i}|i=1,...,N\},\quad \bm{e}^{m}_{i}=m_{i}\odot\bm{e}_{[M]}+(1-m_{i})\odot\bm{e}^{p}_{i},
\end{equation}
where $\delta(\cdot)$ is the indicator function and $m_{i}=\delta(i\in\mathcal{M})$. After that, the masked embeddings $\mathcal{E}_{m}$ will be fed into the Temporal Transformer. The output embeddings $\mathcal{E}^{f}$ will be used to predict the indices of the corresponding codes from the codex in the DPQ through a linear classifier:
\begin{equation}
  p(z^{g}_{i}|\bm{e}^{f}_{i})=\mathrm{softmax}(\mathrm{Linear}(\bm{e}^{f}_{i})),
\end{equation}
The total loss $\mathcal{L}_{f}^{\mathcal{M}}$ for training the MAE model is:
\begin{equation}
  \mathcal{L}_{f}^{\mathcal{M}}=-\sum_{i\in\mathcal{M}}\left[m_{i}\odot \sum_{g=1}^{G}\mathrm{log}\ p(z^{g}_{i}|\bm{e}^{f}_{i})\right].
\end{equation}

\section{Experiments}
\subsection{Dataset}
To validate the effectiveness of our proposed framework, we evaluate BrainStratify on two publicly available sEEG datasets \cite{zheng2025discrete,wang2024brain}, as detailed in Table \ref{table:dataset-info}. Considering the lack of open-source intracranial neural datasets, we follow the experimental design outlined by Du-IN \cite{zheng2025discrete} to collect a well-annotated Chinese word-reading (epidural) ECoG dataset; see Appendix \ref{sec:supp-expr-design} for more details. The ECoG electrodes are positioned epidurally \cite{liu2024reclaiming} -- outside the brain’s dura mater rather than directly on the cortex -- minimizing tissue damage compared to traditional intracranial placements \cite{moses2021neuroprosthesis}.

\begin{table}[h]
  \caption{Overview of intracranial neural datasets used in this work.}
  \label{table:dataset-info}
  \centering
  \tabcolsep=0.01\linewidth
  \begin{tabular}{lcccccccc}
    \toprule
    \multirow{2}{*}{\textbf{Name}} & \multirow{2}{*}{\textbf{Type}} & \multirow{2}{*}{\textbf{Task}} & \multirow{1}{*}{\textbf{\# of}} & \multirow{1}{*}{\textbf{\# of}} & \multirow{1}{*}{\textbf{\# of}} & \multirow{1}{*}{\textbf{Trial}} & \multirow{1}{*}{\textbf{\# of}} & \multirow{1}{*}{\textbf{Total}} \\
    & & & \multirow{1}{*}{\textbf{Subjects}} & \multirow{1}{*}{\textbf{Channels}} & \multirow{1}{*}{\textbf{Trials}} & \multirow{1}{*}{\textbf{Length}} & \multirow{1}{*}{\textbf{Classes}} & \multirow{1}{*}{\textbf{Recordings}} \\
    \midrule
    Du-IN \cite{zheng2025discrete} & sEEG & Read & 12 & 109.75 & $\sim$3k & 3s & 61 & 15 hours \\
    BrainTree \cite{wang2024brain} & sEEG & Movie & 10 & 124.9 & $\sim$2k & 4s & 2 & 5.5 hours \\
    Ours & ECoG & Read & 1 & 128 & $\sim$12k & 2.4s & 62 & 22 hours \\
    \bottomrule
  \end{tabular}
\end{table}

\subsection{Implementation Details}
\label{sec:expr-implementation-details}
\paragraph{Preprocess.}We filter the sEEG/ECoG signals between 0.5Hz and 200Hz to remove low-frequency noise. Then, a notch filter of 50Hz (or 60Hz) is applied to avoid power-line interference. Next, neural signals are resampled to 400Hz and re-referenced \cite{li2018optimal} according to the original setting. Finally, z-score normalization is performed on each channel to standardize data scales across all channels.

\paragraph{Model Configurations.}In the Coarse-DL stage, the raw patches are first transformed into patch embeddings with $d=256$. The following Temporal \& Spatial Transformer both contain a 4-layer Transformer encoder with model dimension $d=256$, inner dimension (FFN) $d_{ff}=1024$, and 8 attention heads. In the Fine-DL stage, the raw patches are transformed into patch embeddings with $d=256$, followed by an 8-layer Transformer encoder with model dimension $d=256$, inner dimension (FFN) $d_{ff}=1024$, and 8 attention heads. See Appendix \ref{sec:supp-model-details} for more details.

\paragraph{Pre-training.}We train the VQ-VAE and MAE models using all recordings from each subject, excluding those reserved for validation and testing in downstream tasks. To enhance the robustness of the learned codex and representations, we use data augmentation described in Appendix \ref{sec:supp-data-augmentation}. For each subject, models are trained on 1 GPU (NVIDIA Tesla V100 32GB) for $\sim$ 8 hours.

\paragraph{Fine-tuning.}We split the task recordings into training, validation, and testing splits with a size roughly proportional to 80\%, 10\%, and 10\%. All experiments are conducted on the same machine with the same set of random seeds. The train/validation/test splits are the same across different models. We use data augmentation described in Appendix \ref{sec:supp-data-augmentation} to make the most of the gathered dataset. Experiments are conducted on one V100 GPU by Python 3.11.7 and PyTorch 2.1.2 + CUDA 12.3. The best models are trained on the training set, selected from the validation set according to accuracy, and finally evaluated on the test set. For model comparison, we report the average and standard error values (of all subjects) on six random seeds to obtain comparable results. For subject-wise evaluation, we report the average and standard deviation values (of each subject) in Appendix \ref{sec:supp-subject-evaluation}.

\subsection{Results on Channel Cluster}
\begin{figure}[h]
  \centering
  \includegraphics[width=\linewidth]{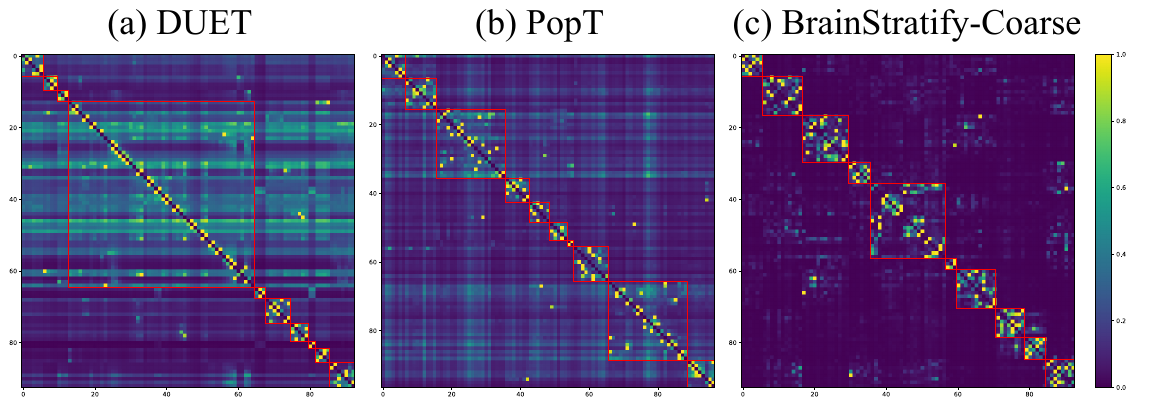}
  \caption{\textbf{The channel connectivity from different methods.}}
  \label{fig:attn-graph}
\end{figure}

Given the sparse distribution of sEEG electrodes across the brain, we begin by identifying functional groups at a coarse level. These groups act as fundamental computational modules \cite{buzsaki2006rhythms,silva2024speech}, wherein channels provide complementary information to encode specific functions (e.g., vocal production, motor control) collectively. After pre-training, we compare the channel connectivity from DUET \cite{qiu2024duet}, PopT \cite{chau2024population}, and our method in Figure \ref{fig:attn-graph}. To standardize comparisons, we normalize connectivity values to a [0,1] range for each channel, accommodating method-specific scaling differences.

Since DUET relies heavily on inter-channel correlations, we pre-train it using raw sEEG signals (without re-referencing), while using re-referenced signals instead even degrades clustering performance. As shown in Figure \ref{fig:attn-graph} (a), DUET struggles to reliably identify functional groups, suggesting that correlation-based metrics may not adequately capture the underlying inter-channel relationships. While PopT employs a spatial context task during pre-training, it faces convergence challenges with limited neural data. Furthermore, when pre-trained across multiple subjects, PopT underperforms our method in identifying functional groups (Figure \ref{fig:attn-graph} (b) \& Table \ref{table:result-chan-select}), primarily due to inherent variability in neural computation across individuals. In contrast, BrainStratify-Coarse effectively captures the sparse inter-channel relationships, aligning with established neuroscientific principles of functional specificity in neural processing \cite{buzsaki2006rhythms,silva2024speech}. Based on these findings, we employ hard clustering ($k=10$) on the estimated channel connectivity from different methods, as illustrated in Figure \ref{fig:attn-graph}.
\begin{table}[h]
  \caption{The 61-word performance of different channel selection strategies on Du-IN dataset \cite{zheng2025discrete}.}
  \label{table:result-chan-select}
  \centering
  \begin{tabular}{l|c|c|cc}
    \toprule
    \multirow{1}{*}{\textbf{Channel Selection}} & \multirow{1}{*}{\textbf{Supervised}} & \multirow{1}{*}{\textbf{\# of Channels (Averaged)}} & \multicolumn{1}{c}{\textbf{Accuracy (\%) $\pm$ ste (\%)}} \\
    \midrule
    SC \cite{wang2023brainbert}    & \Checkmark   & 10 & 48.05$\pm$5.58 \\
    MC \cite{zheng2025discrete}    & \Checkmark   & 10 & \textbf{58.24$\pm$4.39} \\
    \midrule
    CCM \cite{chen2025similarity}  & \XSolidBrush & 109.75 & 29.42$\pm$5.13 \\
    DUET \cite{qiu2024duet}        & \XSolidBrush & 16.08 & 45.14$\pm$3.71 \\
    PopT \cite{chau2024population} & \XSolidBrush & 11.42 & 53.51$\pm$4.27 \\
    \midrule
    BrainStra.-Coarse              & \XSolidBrush & 11.33 & \underline{57.79$\pm$4.16} \\
    \bottomrule
  \end{tabular}
\end{table}

After clustering channels, we select and combine these groups based on their performance in downstream decoding tasks, while different backbones show similar results. We use BrainStratify-Fine (w/o pre-training) as evaluation backbone (Table \ref{table:result-chan-select}). CCM \cite{chen2025similarity} relies on static cluster embeddings, which ignore the dynamic nature of sEEG signals, resulting in unreliable functional group identification. The SC method, which fails to capture complementary information among channels, underperforms the MC strategy. In contrast, BrainStratify-Coarse matches the MC baseline, validating its effectiveness.

\subsection{Results on Neural Decoding}
Table \ref{table:result-main-duin} \& \ref{table:result-main-braintree} compare BrainStratify-Fine against advanced baselines designed for brain signals, with model details in Appendix \ref{sec:supp-baseline-details} \& \ref{sec:supp-model-details}. In Table \ref{table:result-main-duin}, BrainStratify-Fine suppresses all baselines, highlighting DPQ's effectiveness in identifying fine-grained states. H2DiLR \cite{wu2024towards} underperforms Du-IN due to the lack of mask modeling. PopT \cite{chau2024population} employs \texttt{[CLS]} token to aggregate channels, performing between LaBraM \cite{jiang2024large} and EEG-CFMR \cite{song2022eeg}. Due to relatively stable interaction among neuronal populations, temporal models can better handle the variability of neural patterns along the temporal axis by aggregating channels into tokens before modeling temporal relationships, outperforming temporal-spatial models. Notably, this performance gap narrows for (epidural) ECoG data, likely stemming from inherent differences in signal properties across modalities. Besides, BrainStratify-Fine achieves greater improvements over Du-IN \cite{zheng2025discrete} on the (epidural) ECoG dataset compared to the sEEG dataset, likely due to ECoG's lack of spatial resolution enhancement techniques (e.g., bi-polar re-reference).
\begin{table}[h]
  \caption{Results on two word-reading intracranial neural datasets (sEEG \& ECoG). Accuracy (\%) $\pm$ ste (\%) on Du-IN sEEG dataset \cite{zheng2025discrete} and our collected (epidural) ECoG dataset are reported.}
  \label{table:result-main-duin}
  \centering
  \tabcolsep=0.01\linewidth
  \begin{tabular}{l|c|cc|cc}
    \toprule
    \multirow{2}{*}{\textbf{Methods}} & \multirow{2}{*}{\textbf{Chan. Select.}} & \multicolumn{2}{c|}{\textbf{Du-IN \cite{zheng2025discrete}}} & \multicolumn{2}{c}{\textbf{Ours}} \\
    & & Word & Syllable & Word & Syllable \\
    \midrule
    LaBraM \cite{jiang2024large}     & MC & 11.78$\pm$2.70 & - & 28.33$\pm$0.98 & - \\
    CBraMod \cite{wang2024cbramod}   & MC & 11.52$\pm$2.48 & - & 26.88$\pm$1.63 & - \\
    PopT \cite{chau2024population}   & MC & 22.55$\pm$3.26 & - & 29.30$\pm$0.63 & - \\
    \midrule
    EEG-CFMR \cite{song2022eeg}      & MC & 45.82$\pm$4.66 & 62.63$\pm$3.77 & 42.16$\pm$2.46 & 59.59$\pm$0.42 \\
    Du-IN \cite{zheng2025discrete}   & MC & 62.70$\pm$4.69 & 70.66$\pm$3.74 & \underline{52.63$\pm$1.68} & \underline{66.75$\pm$0.72} \\
    H2DiLR \cite{wu2024towards}      & MC & 25.84$\pm$3.12 & 43.29$\pm$1.84 & 32.21$\pm$1.33 & 46.63$\pm$0.48 \\
    \midrule
    BrainStra.-Fine                 & MC              & \underline{66.35$\pm$3.86} & \textbf{75.54$\pm$3.19} & \textbf{58.50$\pm$1.51} & \textbf{70.66$\pm$0.76} \\
    BrainStra.-Fine                 & BrainStra.-Coarse & \textbf{66.44$\pm$3.65} & \underline{75.36$\pm$3.17} & - & - \\
    \bottomrule
  \end{tabular}
\end{table}

As shown in Table \ref{table:result-main-braintree}, BrainStratify-Coarse identifies functional groups that effectively capture complementary neural information related to movie tasks, enhancing downstream channel aggregation. When evaluated on channels selected by BrainStratify-Coarse, most models outperform the PopT baseline \cite{chau2024population}, which uses single-channel (SC) selection. Besides, BrainStratify-Fine surpasses all advanced baselines, further demonstrating the superiority of temporal modeling approaches.
\begin{table}[h]
  \caption{Results on Brain Treebank sEEG dataset \cite{wang2024brain}.}
  \label{table:result-main-braintree}
  \centering
  \begin{tabular}{l|c|cccc}
    \toprule
    \multirow{2}{*}{\textbf{Methods}} & \multirow{2}{*}{\textbf{Chan. Select.}} & \multicolumn{4}{c}{\textbf{ROC-AUC $\pm$ ste}} \\
    & & Pitch & Volumn & Sent. Onset & Word Onset \\
    \midrule
    PopT \cite{chau2024population}   & SC & 0.74$\pm$0.03 & 0.87$\pm$0.03 & 0.90$\pm$0.01 & 0.93$\pm$0.02 \\
    \midrule
    LaBraM \cite{jiang2024large}     & BrainStra.-Coarse & 0.72$\pm$0.02 & 0.87$\pm$0.02 & 0.90$\pm$0.02 & 0.94$\pm$0.01 \\
    CBraMod \cite{wang2024cbramod}   & BrainStra.-Coarse & 0.70$\pm$0.02 & 0.84$\pm$0.03 & 0.90$\pm$0.02 & 0.94$\pm$0.01 \\
    PopT \cite{chau2024population}   & BrainStra.-Coarse & 0.74$\pm$0.02 & 0.88$\pm$0.02 & \underline{0.94$\pm$0.01} & 0.96$\pm$0.01 \\
    \midrule
    EEG-CFMR \cite{song2022eeg}      & BrainStra.-Coarse & 0.77$\pm$0.02 & 0.89$\pm$0.02 & 0.92$\pm$0.02 & 0.94$\pm$0.01 \\
    Du-IN \cite{zheng2025discrete}   & BrainStra.-Coarse & \underline{0.78$\pm$0.02} & \underline{0.90$\pm$0.02} & 0.94$\pm$0.01 & \underline{0.97$\pm$0.01} \\
    H2DiLR \cite{wu2024towards}      & BrainStra.-Coarse & 0.77$\pm$0.02 & 0.88$\pm$0.02 & 0.88$\pm$0.02 & 0.90$\pm$0.01 \\
    \midrule
    BrainStra.-Fine                 & BrainStra.-Coarse & \textbf{0.79$\pm$0.02} & \textbf{0.91$\pm$0.02} & \textbf{0.95$\pm$0.02} & \textbf{0.98$\pm$0.01} \\
    \bottomrule
  \end{tabular}
\end{table}

\subsection{Ablation Study}
\begin{figure}[h]
  \centering
  \includegraphics[width=\linewidth]{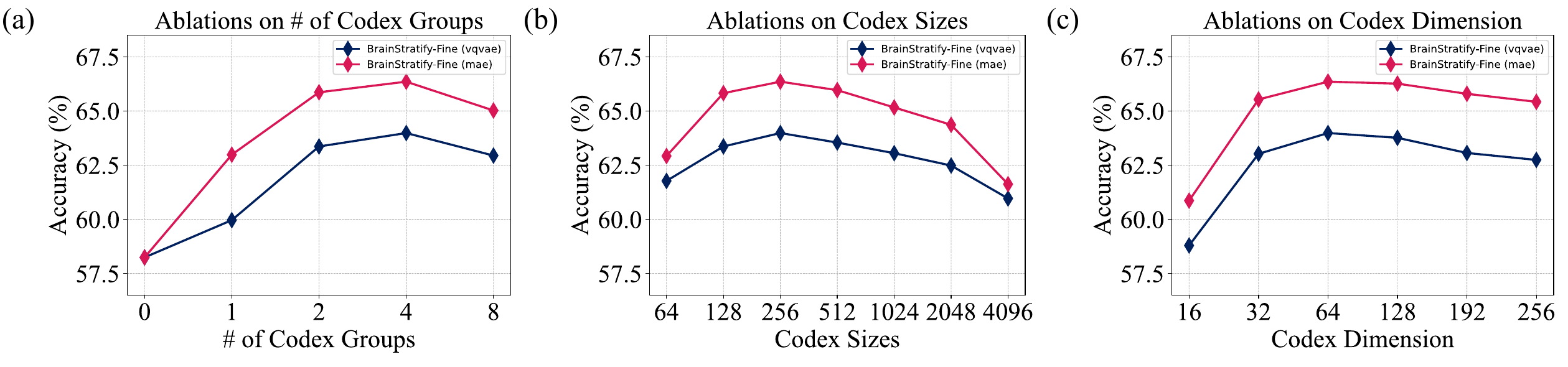}
  \caption{\textbf{Ablation study on different codex groups, codex sizes, and codex dimensions.} We report the 61-word performance on Du-IN dataset \cite{zheng2025discrete}; see Appendix \ref{sec:supp-model-details} for details of model variants.}
  \label{fig:ablation-codex}
\end{figure}

We conduct thorough ablation studies, with additional analyses provided in Appendix \ref{sec:supp-ablation-study}.
\paragraph{Discrete Codex.}We evaluate the hyper-parameters of DPQ through comprehensive ablation studies. Figure \ref{fig:ablation-codex} shows BrainStratify-Fine's 61-word classification performance on the Du-IN dataset across different initialization strategies (details in Appendix \ref{sec:supp-model-details}). We evaluate performance against varying codex groups (from 0 to 8) to ascertain if the number of codex groups affects the quality of the learned codex. To maintain the capacity of the codex with $G=1$, we set $N_{codex}$ to 2048, while for the other settings, $N_{codex}$ is set to 256. As illustrated in Figure \ref{fig:ablation-codex} (a), since the channels are pre-selected based on target functional groups, even a small number of codex groups (e.g., $G=4$) can effectively decouple distinct neural dynamics \cite{chapeton2022micro,mudrik2024sibblings,mudrik2025creimbo}. We also assess performance across different codex sizes (from 64 to 4096) to ascertain if codex size affects the quality of the learned codex. As illustrated in Figure \ref{fig:ablation-codex} (b), while extremely small codex size lacks representation diversity, extremely large codex size often leads to codex collapse. We suspect that our existing training data might not be adequate for larger codex sizes. Furthermore, our experiments suggest that the model performs best when the dimension of module code $\bm{\mathrm{z}}_{q[g]}(\bm{e}^{f}_{i})$, denoted as $d_{codex}=64$, is slightly smaller than the model dimension, $d=256$, resulting in more effective regularization.

\section{Limitations}
\label{sec:limitations}
Given the sparse distribution of sEEG electrodes across the brain, BrainStratify is currently limited to identifying functional modules rather than full functional networks. Nevertheless, these modules serve as fundamental building blocks for such networks. Extending our framework to infer dynamic network topology would enrich its neuroscientific insight if higher-density sEEG electrode configurations become available and models are refined to capture time-lagged spatial relationships.

Due to the scarcity of open-source intracranial neural datasets, BrainStratify is currently evaluated on three datasets (e.g., vocal production, speech perception). While this scope is limited, the framework’s design affords clear interpretability grounded in brain organization, ensuring a solid neuroscience foundation \cite{buzsaki2006rhythms,silva2024speech,chapeton2022micro,mudrik2025creimbo}. We anticipate that BrainStratify can be extended to decode other cognitive states (e.g., motor control, image perception) when more datasets become publicly available.

\section{Conclusion}
This paper proposes BrainStratify, a novel framework that disentangles intracranial neural dynamics in a Coarse-to-Fine way to identify fine-grained states. Comprehensive experiments demonstrate that BrainStratify-Coarse reliably identifies functional groups from sEEG signals, surpassing existing channel clustering baselines. In addition, BrainStratify-Fine effectively decouples distinct neural dynamics within target functional groups, achieving superior performance across intracranial neural modalities (e.g., sEEG, ECoG) compared to advanced neural decoding baselines. Overall, our approach -- inspired by neuroscience findings -- is suitable for decoding speech from intracranial neural signals, advancing toward clinically viable and transparent neuroprosthetic systems.

\clearpage
\subsection*{Reproducibility Statement}
Code to train models and reproduce the results is submitted as part of the supplementary materials and can be accessed here: \href{TODO}{TODO}.

\subsubsection*{Ethics Statement}
\label{sec:ethics-statement}
Experiments that contribute to this work were approved by IRB. All subjects consent to participate. All electrode locations are exclusively dictated by clinical considerations.

Our informed consent signing process is as follows:
\begin{enumerate}
  \item If the experimental participants are adults and have full civil capacity, we will ask them to sign a written informed consent after the participants have fully informed consent;
  \item If the experimental participants are minors or do not have full civil capacity, we will ask the participant's legal guardian to sign a written informed consent after the participants and their legal guardians have fully informed consent.
\end{enumerate}

Our informed consent form includes the following points:
\begin{enumerate}
  \item Contact information of research institutions and researchers;
  \item Research direction and purpose;
  \item Risks involved in the research;
  \item Personal information, data and usage methods to be used in the research;
  \item Privacy protection statement (all personal identification information (PII) will not be disclosed);
  \item Data storage statement (retained after deleting all personal identification information (PII));
  \item Voluntary statement of participants;
  \item Statement that participants can withdraw unconditionally at any time.
\end{enumerate}

Our data storage and protection procedures include the following processes:
\begin{enumerate}
  \item Our data collection, transfer, and analysis tasks are only completed by researchers who have signed relevant confidentiality agreements;
  \item The collected raw data will be copied twice as soon as possible, one copy to a storage computer that is not connected to the Internet and encrypted, and the other copy to a mobile hard disk and encrypted and stored offline;
  \item The use of the data is only authorized to the research leader and the main researchers (less than 5 people), among which the main researchers can only access data that does not contain personal identification information (PII);
  \item After the study is completed, all personal identification information (PII) on both nodes (storage computer, mobile hard disk) will be deleted immediately.
\end{enumerate}

To prevent unauthorized access or possible data leakage, we use double encryption on the storage computer, that is, a static password and a dynamic password (received by mobile phone or email); physical isolation is used on the mobile hard disk, that is, it is locked in a filing cabinet, and the key is only kept by the research leader and the main researchers.

\bibliography{neurips_2025}
\bibliographystyle{plain}

\clearpage
\appendix

\section{Experiment Design}
\label{sec:supp-expr-design}
Due to the lack of open-source intracranial neural datasets related to speech, we follow the experimental design outlined in Du-IN \cite{zheng2025discrete} to collect a well-annotated Chinese word-reading (epidural) ECoG dataset, including one subject (female; aged 67) who has lost her ability to communicate or perform daily tasks due to amyotrophic lateral sclerosis (ALS).

\begin{figure}[h]
  \centering
  \includegraphics[width=\linewidth]{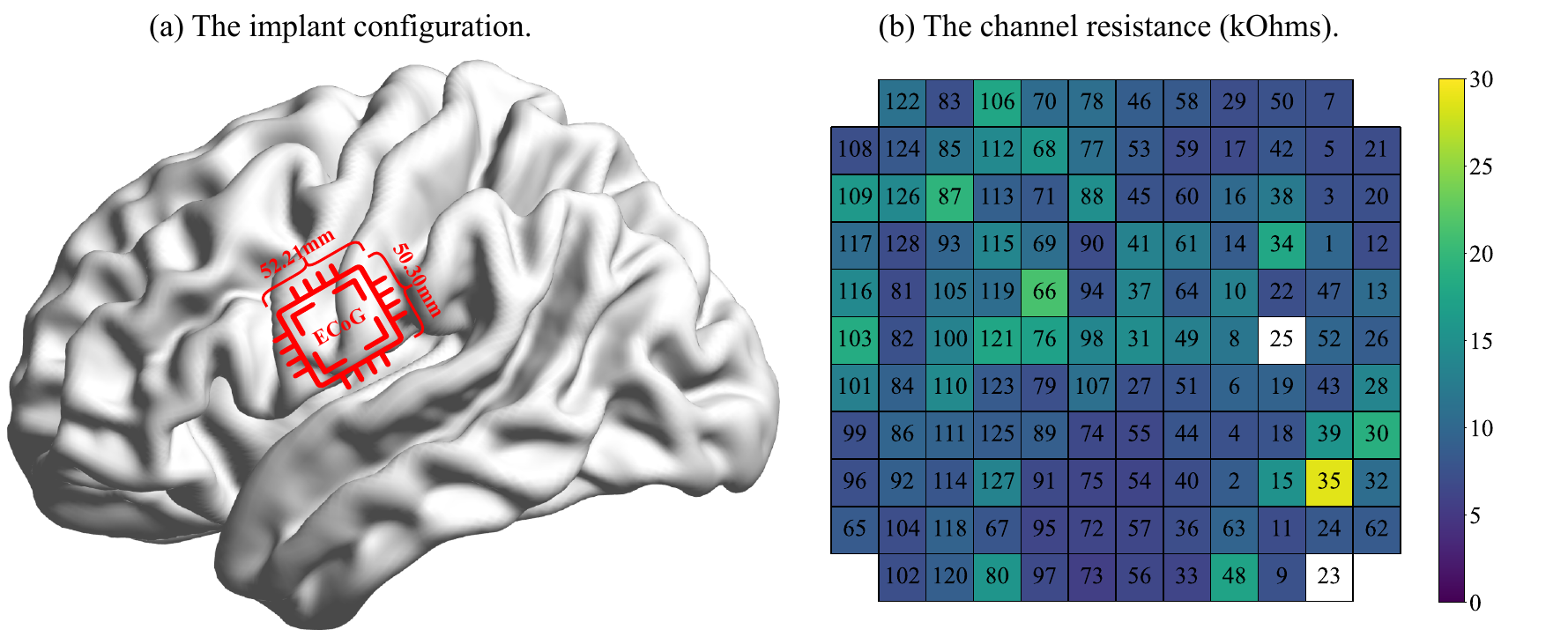}
  \caption{\textbf{Overview of ECoG configuration.} \textbf{(a).} The implant configuration. Our developed (epidural) ECoG is placed above vSMC, which is involved in vocal production \cite{silva2024speech}. \textbf{(b).} The channel resistance. Electrodes at the four corners are excluded for downstream analysis.}
  \label{fig:ecog-grid}
\end{figure}

We developed a minimally invasive BCI \cite{branco2023nine,liu2024reclaiming} with an 11$\times$12 (epidural) ECoG grid (50.30mm $\times$ 52.21mm) above the ventral sensorimotor cortex (vSMC) to restore the speech functions of that subject, as shown in Figure \ref{fig:ecog-grid} (a). With wireless powering and neural data transmission, this system enables real-time speech neuroprosthesis in home use. After excluding the four corner electrodes, we analyzed neural recordings from the remaining 128 channels, as shown in Figure \ref{fig:ecog-grid} (b). All electrodes (except No.23 and No.25) exhibit consistently low impedance levels during neural recordings, ensuring robust signal fidelity.

In the word-reading task, the subject attempts to speak individual words from a 62-word set while we record her brain activity (measured by ECoG). A list of the words contained in this 62-word set is provided in Table \ref{table:word-set}.

All data are collected as a series of "blocks" (42 blocks in total), with each block lasting about 20 minutes and consisting of multiple trials. During each block of this task, all words (from the 62-word set) are presented individually 5 times, leading to a total of 310 trials.

Each trial in a block of this task starts with one word shown on the screen in white text. After 0.2 seconds, the text will turn green and remain on the screen for 2.2 seconds. This color transition from white to green represents the go cue for each trial, and the subject is instructed to speak the word aloud as soon as the text turns green. Afterward, the text will be replaced with a blank screen with a centered cross. After 0.8 seconds, the task continues to the next trial. The word presentation order is randomized within each task block.

Besides, we also collected non-task recordings of subjects in their daily life. There are roughly 8 hours of non-task recordings during wakefulness. In summary, for each subject, we collect about 22 hours of sEEG recordings, of which 14 hours are task recordings.

\begin{table}[h]
  \caption{The Chinese words and their corresponding English translations in the 62-word set.}
  \label{table:word-set}
  \centering
  \begin{tabular}{cc|cc|cc}
    \toprule
    \textbf{Words} & \textbf{Translations} & \textbf{Words} & \textbf{Translations} & \textbf{Words} & \textbf{Translations} \\
    \midrule
    \begin{CJK}{UTF8}{gbsn}的\end{CJK} & of & \begin{CJK}{UTF8}{gbsn}对\end{CJK} & right & \begin{CJK}{UTF8}{gbsn}在\end{CJK} & exist \\
    \midrule
    \begin{CJK}{UTF8}{gbsn}把\end{CJK} & handle & \begin{CJK}{UTF8}{gbsn}是\end{CJK} & be & \begin{CJK}{UTF8}{gbsn}要\end{CJK} & want \\
    \midrule
    \begin{CJK}{UTF8}{gbsn}和\end{CJK} & and & \begin{CJK}{UTF8}{gbsn}你\end{CJK} & you & \begin{CJK}{UTF8}{gbsn}这\end{CJK} & this \\
    \midrule
    \begin{CJK}{UTF8}{gbsn}去\end{CJK} & go & \begin{CJK}{UTF8}{gbsn}有\end{CJK} & have & \begin{CJK}{UTF8}{gbsn}没\end{CJK} & without \\
    \midrule
    \begin{CJK}{UTF8}{gbsn}他\end{CJK} & he & \begin{CJK}{UTF8}{gbsn}看\end{CJK} & look & \begin{CJK}{UTF8}{gbsn}我\end{CJK} & I \\
    \midrule
    \begin{CJK}{UTF8}{gbsn}给\end{CJK} & give & \begin{CJK}{UTF8}{gbsn}不\end{CJK} & no & \begin{CJK}{UTF8}{gbsn}都\end{CJK} & all \\
    \midrule
    \begin{CJK}{UTF8}{gbsn}就\end{CJK} & at once & \begin{CJK}{UTF8}{gbsn}帮\end{CJK} & help & \begin{CJK}{UTF8}{gbsn}好\end{CJK} & good \\
    \midrule
    \begin{CJK}{UTF8}{gbsn}找\end{CJK} & find & \begin{CJK}{UTF8}{gbsn}陪\end{CJK} & accompany & \begin{CJK}{UTF8}{gbsn}热\end{CJK} & hot \\
    \midrule
    \begin{CJK}{UTF8}{gbsn}冷\end{CJK} & cold & \begin{CJK}{UTF8}{gbsn}人\end{CJK} & people & \begin{CJK}{UTF8}{gbsn}想\end{CJK} & think \\
    \midrule
    \begin{CJK}{UTF8}{gbsn}吗\end{CJK} & ? & \begin{CJK}{UTF8}{gbsn}出\end{CJK} & out & \begin{CJK}{UTF8}{gbsn}医生\end{CJK} & doctor \\
    \midrule
    \begin{CJK}{UTF8}{gbsn}可以\end{CJK} & can & \begin{CJK}{UTF8}{gbsn}睡觉\end{CJK} & sleep & \begin{CJK}{UTF8}{gbsn}说话\end{CJK} & speak \\
    \midrule
    \begin{CJK}{UTF8}{gbsn}休息\end{CJK} & rest & \begin{CJK}{UTF8}{gbsn}问题\end{CJK} & question & \begin{CJK}{UTF8}{gbsn}家人\end{CJK} & family \\
    \midrule
    \begin{CJK}{UTF8}{gbsn}谢谢\end{CJK} & thanks & \begin{CJK}{UTF8}{gbsn}朋友\end{CJK} & friend & \begin{CJK}{UTF8}{gbsn}吃饭\end{CJK} & eat \\
    \midrule
    \begin{CJK}{UTF8}{gbsn}手机\end{CJK} & cell phone & \begin{CJK}{UTF8}{gbsn}喝水\end{CJK} & drink water & \begin{CJK}{UTF8}{gbsn}心情\end{CJK} & mood \\
    \midrule
    \begin{CJK}{UTF8}{gbsn}厕所\end{CJK} & toilet & \begin{CJK}{UTF8}{gbsn}快乐\end{CJK} & happy & \begin{CJK}{UTF8}{gbsn}困难\end{CJK} & difficulty \\
    \midrule
    \begin{CJK}{UTF8}{gbsn}紧急\end{CJK} & urgent & \begin{CJK}{UTF8}{gbsn}护士\end{CJK} & nurse & \begin{CJK}{UTF8}{gbsn}感觉\end{CJK} & feel \\
    \midrule
    \begin{CJK}{UTF8}{gbsn}舒服\end{CJK} & comfortable & \begin{CJK}{UTF8}{gbsn}电脑\end{CJK} & computer & \begin{CJK}{UTF8}{gbsn}坐着\end{CJK} & sitting \\
    \midrule
    \begin{CJK}{UTF8}{gbsn}躺下\end{CJK} & lie down & \begin{CJK}{UTF8}{gbsn}洗澡\end{CJK} & bath & \begin{CJK}{UTF8}{gbsn}现在\end{CJK} & now \\
    \midrule
    \begin{CJK}{UTF8}{gbsn}书籍\end{CJK} & book & \begin{CJK}{UTF8}{gbsn}医院\end{CJK} & hospital & \begin{CJK}{UTF8}{gbsn}衣服\end{CJK} & clothes \\
    \midrule
    \begin{CJK}{UTF8}{gbsn}一起\end{CJK} & together & \begin{CJK}{UTF8}{gbsn}散步\end{CJK} & walk & \begin{CJK}{UTF8}{gbsn}呼吸\end{CJK} & breathe \\
    \midrule
    \begin{CJK}{UTF8}{gbsn}非常\end{CJK} & very & \begin{CJK}{UTF8}{gbsn}回家\end{CJK} & go home & - & - \\
    \bottomrule
  \end{tabular}
\end{table}

\clearpage
\section{Task Details}
\label{sec:supp-task-details}
In experiments, we evaluate multiple neural decoding tasks on two publicly available sEEG datasets \cite{zheng2025discrete,wang2024brain} and our collected word-reading (epidural) ECoG dataset.

\subsection{Du-IN sEEG dataset}
We follow the same task specification in Du-IN \cite{zheng2025discrete}. Each trial contains 3-second neural activity.

\paragraph{Word Reading.}The subject speaks aloud individual words from a 61-word set while his neural activity and voice are simultaneously recorded. Labels are balanced in the experiment design.

We follow the evaluation protocol in Du-IN \cite{zheng2025discrete}. Since this task is a classification (CLS) task, we flatten embeddings and add a linear head after either pre-trained or randomly initialized models. Training employs cross-entropy loss, with results quantified using top-1 accuracy scores.

Besides, following Willett et al. \cite{willett2023high,fan2023plug}, we evaluate the 49-syllable connectionist temporal classification (CTC) task. Considering the difference between English and Chinese, we utilize syllables from Pinyin \cite{wang1973chinese}, a widely adopted phonetic representation system based on the Latin alphabet, as basic units that have the potential to support open-set speech decoding tasks. The set of 49 syllables includes:
\begin{itemize}
  \item 1 CTC blank token (i.e., "-"),
  \item 1 silence token (i.e., "$|$"),
  \item 23 initial syllable tokens,
  \item 24 final syllable tokens.
\end{itemize}
Take the pre-determined word "\begin{CJK}{UTF8}{gbsn}电脑\end{CJK}" (i.e., "computer") for example, the corresponding syllable label sequence $\bm{y}$ is ["$|$", "d", "i", "an", "$|$", "n", "ao", "$|$"]. Since this task is a connectionist temporal classification (CTC) task, we add a linear head after either pre-trained or randomly initialized models. Training employs connectionist temporal classification (CTC) loss, with results quantified using syllable error rate scores. Specifically, the syllable error rate (SER) is computed based on the syllable edit distances, which is widely adopted in neural decoding research \cite{fan2023plug,willett2023high,metzger2023high}. To align with the trend of top-1 accuracy, we present results as \texttt{(1-SER)} rather than raw \texttt{SER} values.

\subsection{Brain Treebank sEEG dataset}
While adopting the same task specification in PopT \cite{chau2024population}, we narrowed the analysis window to [-2.0,2.0]s around word onset (vs. PopT's [-2.5,2.5]s), yielding 4-second neural activity per trial.

\paragraph{Pitch.}The pitch of a given word is extracted using Librosa's \texttt{piptrack} function over a Mel-spectrogram (sampling rate 48,000 Hz, FFT window length of 2048, hop length of 512, and 128 mel filters). For this task, for a given session, the positive examples consist of words in the top quartile of pitch, and the negative examples are the words in the bottom quartile.

\paragraph{Volume.}The volume of a given word is computed as the average intensity of root-mean-square (RMS) (\texttt{rms} function, frame and hop lengths 2048 and 512 respectively). For this task, for a given session, the positive examples are the words in the top quartile of volume, and the negative examples are the words in the bottom quartile.

\paragraph{Sent. Onset (Sentence Onset).}The negative examples are intervals of activity from 1s periods during which no speech occurs in the movie. The positive examples are intervals of brain activity that correspond with hearing the first word of a sentence.

\paragraph{Word Onset (Speech vs. Non-speech).}The negative examples are intervals of activity from 1s periods during which no speech occurs in the movie. The positive examples are intervals of brain activity that correspond with dialogue being spoken in the stimuli movie.

For each task, we follow the evaluation protocol in PopT \cite{chau2024population}, using the specified movie for downstream classification. Since these tasks are binary classification (CLS) tasks, we flatten embeddings and add a linear head after either pre-trained or randomly initialized models. Training employs binary cross-entropy (BCE) loss, with results quantified using ROC-AUC scores.

\subsection{Our (epidural) ECoG dataset}
The experiment design is discussed in Appendix \ref{sec:supp-expr-design}. Each trial contains 2.4-second neural activity.

\paragraph{Word Reading.}The subject attempts to speak individual words from a 62-word set while his neural activity is recorded. Labels are balanced in the experiment design.

We follow the evaluation protocol in Du-IN \cite{zheng2025discrete}. Since this task is a classification (CLS) task, we flatten embeddings and add a linear head after either pre-trained or randomly initialized models. Training employs cross-entropy loss, with results quantified using top-1 accuracy scores.

Besides, following the aforementioned evaluation protocol for the Du-IN sEEG dataset, we evaluate the 49-syllable connectionist temporal classification (CTC) task. Training employs connectionist temporal classification (CTC) loss, with results quantified using syllable error rate scores.

\clearpage
\section{Baseline Details}
\label{sec:supp-baseline-details}
\subsection{Channel Cluster Baselines}
In experiments, we compare our model to the advanced channel cluster methods \cite{chen2025similarity,qiu2024duet} in time series analysis. The details of these baseline models are given here:
\begin{itemize}
  \item \textbf{CCM} \cite{chen2025similarity}: A time-series forecasting model that learns channel clustering based on intrinsic similarities and creates prototype embeddings for each cluster via a cross-attention mechanism. Since sEEG is a unique type of time series, this model is suitable to serve as a baseline for comparison.
  \item \textbf{DUET} \cite{qiu2024duet}: A time-series forecasting model that captures the relationships among channels in the frequency domain through metric learning and applies sparsification to mitigate the adverse effects of noisy channels. Since sEEG is a unique type of time series, this model is suitable to serve as a baseline for comparison.
\end{itemize}
The detailed implementations of these baseline models are given here:
\begin{itemize}
  \item As the CCM method \cite{chen2025similarity} relies heavily on the correlation-based similarities among channels, we use raw sEEG signals of each subject (before bi-polar re-reference \cite{li2018optimal}) as inputs. Bi-polar re-reference disrupts these correlations, causing CCM to fail convergence reliably. Because CCM does not generate an intermediate inter-channel similarity matrix (i.e., $\mathcal{P}$ in this work), we directly utilize the assigned clusters for downstream channel selection.
  \item Like CCM, the DUET method \cite{qiu2024duet} relies on raw inter-channel correlations. We use raw sEEG signals of each subject (before bi-polar re-reference) as inputs, ensuring stable model convergence. In practice, DUET produces an intermediate inter-channel similarity matrix (i.e., $\mathcal{P}$ in this work) and integrates this matrix via a masked attention mechanism, enhancing time series forecasting. Therefore, we directly apply spectral cluster \cite{ng2001spectral} on this matrix to derive channel clusters for downstream channel selection.
\end{itemize}
Although PopT \cite{chau2024population} was originally proposed as an sEEG foundation model, we also compare our model with it. PopT introduces channel connectivity based on the pre-trained model, which provides an alternative to conventional coherence analysis. Similar to our model, we use bi-polar (or laplacian) re-referenced sEEG signals as inputs. Since PopT struggles to converge with limited data, we pre-train PopT with sEEG signals from all available subjects, which is the original setting \cite{chau2024population}. After computing channel connectivity based on the pre-trained model, we apply spectral cluster \cite{ng2001spectral} on this matrix to derive channel clusters for downstream channel selection.
\subsection{Neural Decoding Baselines}
In experiments, we compare our model to the existing supervised or self-supervised neural decoding methods \cite{jiang2024large,wang2024cbramod,chau2024population,song2022eeg,zheng2025discrete,wu2024towards} on brain signals. The details of these baseline models are given here:
\begin{itemize}
  \item \textbf{LaBraM} \cite{jiang2024large}: A self-supervised model for EEG recordings that learns generic representations with tremendous EEG data. LaBraM models temporal and spatial dependencies simultaneously, serving as an EEG foundation model. Since the spatial embeddings are originally pre-defined according to the EEG caps, we replace the learnable spatial embeddings with hard-coded spatial embeddings from PopT \cite{chau2024population} to enable multi-subject pre-training under the sEEG setting. Since the data modes of EEG and sEEG are similar, this model is suitable to serve as a baseline for comparison.
  \item \textbf{CBraMod} \cite{wang2024cbramod}: A self-supervised model for EEG recordings that captures the heterogeneity between temporal and spatial dependencies. CBraMod combines a criss-cross attention mechanism with asymmetric conditional positional encoding (ACPE) to effectively model temporal-spatial dependencies among EEG patches. CBraMod serves as an EEG foundation model, achieving SOTA performance on various EEG tasks. Since the data modes of EEG and sEEG are similar, this model is suitable to serve as a baseline for comparison.
  \item \textbf{PopT} \cite{chau2024population}: A self-supervised model for sEEG that learns population-level codes for arbitrary ensembles of neural recordings at scale. PopT stacks on top of pre-trained temporal embeddings and enhances downstream decoding by enabling the learned aggregation of multiple spatially sparse channels. PopT serves as an sEEG foundation model, achieving SOTA performance on Brain Treebank \cite{wang2024brain}. As a foundation model in the sEEG pre-training field, this model is suitable to serve as a baseline for comparison.
  \item \textbf{EEG-CFMR} \cite{song2022eeg}: A supervised model for EEG that consists of both CNN module and Transformer module, to encapsulate local and global features in a unified EEG classification framework. EEG-CFMR is mainly designed for EEG-based motor imagination tasks. Since the data modes of EEG and sEEG are similar, this model is suitable to serve as a baseline for comparison.
  \item \textbf{Du-IN} \cite{zheng2025discrete}: A self-supervised model for sEEG-based speech decoding that learns contextual embeddings based on region-level tokens through discrete codex-guided mask modeling. Du-IN achieves SOTA performance on sEEG-based speech decoding using the Du-IN dataset \cite{zheng2025discrete}. As a strong baseline in sEEG-based speech decoding, this model is suitable to serve as a baseline for comparison.
  \item \textbf{H2DiLR} \cite{wu2024towards}: A self-supervised model for sEEG-based tone decoding that disentangles and learns both the homogeneity and heterogeneity from intracranial sEEG recordings across multiple subjects. H2DiLR achieves SOTA performance on sEEG-based tone decoding using the sEEG dataset from Feng et al. \cite{feng2023acoustic}. As a strong baseline in sEEG-based tone decoding (part of speech decoding), this model is suitable to serve as a baseline for comparison.
\end{itemize}
The detailed implementations of these baseline models are given here:
\begin{itemize}
  \item For the LaBraM method \cite{jiang2024large}, the hyper-parameters are the same as the original implementation of the LaBraM-Base model. In practice, foundation models are pre-trained on massive neural datasets. Therefore, their architectures remain fixed after the pre-training stage, limiting post-hoc modifications. The data samples are resampled to the specified sampling rate (i.e., 200 Hz).
  \item For the CBraMod method \cite{wang2024cbramod}, the hyper-parameters are the same as the original implementation of the CBraMod model. The data samples are resampled to the specified sampling rate (i.e., 200 Hz).
  \item For the PopT method \cite{chau2024population}, the hyper-parameters are the same as the original implementation of the PopT model. The data samples are resampled to the specified sampling rate (i.e., 2048 Hz).
  \item For the EEG-CFMR method \cite{song2022eeg}, the hyper-parameters are the same as the original implementation of the EEG-CFMR model. The data samples are resampled to the specified sampling rate (i.e., 250 Hz).
  \item For the Du-IN method \cite{zheng2025discrete}, the hyper-parameters are the same as the original implementation of the Du-IN model. The data samples are resampled to the specified sampling rate (i.e., 1000 Hz).
  \item For the H2DiLR method \cite{wu2024towards}, the hyper-parameters are the same as the original implementation of the H2DiLR model. The data samples are resampled to the specified sampling rate (i.e., 1000 Hz).
\end{itemize}
When evaluating the decoding performance of these baseline models, we follow the same experiment setup of our model; see Appendix \ref{sec:supp-model-details} and Appendix \ref{sec:supp-data-augmentation} for more details.

For the self-supervised methods, the pre-training setup follows the original setup of each model:
\begin{itemize}
  \item For the LaBraM model, we include neural recordings from all available subjects within each dataset for pre-training. The data samples are 4 seconds.
  \item For the CBraMod model, we include neural recordings from all available subjects within each dataset for pre-training. The data samples are 4 seconds.
  \item For the PopT model, we include neural recordings from all available subjects within each dataset for pre-training. The data samples are 4 seconds.
  \item For the Du-IN model, we include neural recordings from each subject for pre-training. The data samples are 4 seconds.
  \item For the H2DiLR model, we include neural recordings from all available subjects within each dataset for pre-training. The data samples are 4 seconds.
\end{itemize}

\clearpage
\section{Model Details}
\label{sec:supp-model-details}
\subsection{BrainStratify-Coarse}
The BrainStratify-Coarse (BrainStratify-C) model is a general architecture for sEEG-based functional group identification, as shown in Figure \ref{fig:cfd} (a). The architecture of BrainStratify-Coarse contains three parts: (1) Patch Tokenizer, (2) Temporal \& Spatial Transformer, and (3) Channel Cluster Module. During the pre-training stage, one additional "Token Classification (CLS) Head" is added after the "Spatial Transformer" for spatial context classification. The hyperparameters for BrainStratify-Coarse training are shown in Table \ref{table:model-brainstratify-coarse}.
\paragraph{Spatial Context Task.}Since sEEG channels capture local and depth information from different brain regions, their recordings inherently capture unique neural information with minimal overlap. This makes the spatial context task better suited for learning inter-channel relationships compared to mask-based reconstruction approaches \cite{jiang2024large,wang2024cbramod}. To ensure balanced label distribution, we designate only 10\% of unreplaced channels as positive samples during pre-training. For all subjects used in this work, the model converges rapidly to $\geq$ 95\%.
\paragraph{Channel Cluster Module.}After pre-training with the spatial context task, we calculate the channel connectivity $\mathcal{P}\in\mathbb{R}^{C\times C}$ following Algorithm \ref{algo:channel-connectivity}. Then, spectral cluster \cite{ng2001spectral} is applied to group channels into functional clusters, using scikit-learn’s \cite{pedregosa2011scikit} \texttt{cluster.SpectralClustering} with default function arguments.
\begin{table}[h]
  \caption{The hyperparameters for BrainStratify-Coarse training.}
  \label{table:model-brainstratify-coarse}
  \centering
  \begin{tabular}{ccc}
    \toprule
    \textbf{Module} & \textbf{Name} & \textbf{Value} \\
    \midrule
    \multirow{6}{*}{Patch Tokenizer} & \# of Input Channels & \{1,128,128\} \\
    & \# of Output Channels & \{128,128,128\} \\
    & Kernel Size & \{9,9,3\} \\
    & Stride & \{5,5,2\} \\
    & Padding & \{4,4,1\} \\
    & Flatten Window & 2 \\
    \midrule
    \multirow{7}{*}{Temporal Transformer} & \# of Transformer Layers & 4 \\
    & Hidden Size & 256 \\
    & MLP Size & 1024 \\
    & MLP Dropout Ratio & \{0.2,0.\} \\
    & \# of Attention Heads & 8 \\
    & Attention Head Size & 64 \\
    & Attention Dropout Ratio & 0.2 \\
    \midrule
    \multirow{7}{*}{Spatial Transformer} & \# of Transformer Layers & 4 \\
    & Hidden Size & 256 \\
    & MLP Size & 1024 \\
    & MLP Dropout Ratio & \{0.2,0.\} \\
    & \# of Attention Heads & 8 \\
    & Attention Head Size & 64 \\
    & Attention Dropout Ratio & 0.2 \\
    \midrule
    \multirow{1}{*}{Token CLS Head} & Linear Projection & $256\rightarrow 1$ \\
    \midrule
    \multirow{9}{*}{\makecell[c]{Optimizer}} & Batch Size & 32 \\
    & Maximum Learning Rate & 3e-4 \\
    & Minimum Learning Rate & 5e-6 \\
    & Learning Rate Scheduler & Cosine \\
    & Optimizer Type & AdamW \\
    & Adam $\beta$ & $(0.9,0.99)$ \\
    & Weight Decay & 0.05 \\
    & Total Epochs & 100 \\
    & Warm-up Epochs & 10 \\
    \bottomrule
  \end{tabular}
\end{table}
\begin{algorithm}
  \caption{The calculation of channel connectivity $\mathcal{P}\in\mathbb{R}^{C\times C}$.}
  \label{algo:channel-connectivity}
  \begin{algorithmic}
  \Require $\{\mathcal{X}_{i}\in \mathbb{R}^{C \times T}|i=1,...,N_{\mathrm{samples}}\}$ \Comment{$N_{\mathrm{samples}}$ is the number of samples.}
  \State $\mathcal{P} \gets \mathbf{0}_{C\times C}$ \Comment{$\mathcal{P}\in \mathbb{R}^{C\times C}$ is initialized as 0s.}
  \While{$i \leq N_{\mathrm{samples}}$}
  \State $\hat{\mathcal{P}} \gets \mathrm{model}(\mathcal{X}_{i})$ \Comment{$\hat{\mathcal{P}}\in \mathbb{R}^{N_{\mathrm{layer}}\times N_{\mathrm{head}}\times C\times C}$ is spatial attention scores.}
  \State $\hat{\mathcal{P}} \gets \mathrm{mean}(\hat{\mathcal{P}}, \mathrm{axes=[0,1]})$ \Comment{$\hat{\mathcal{P}}\in \mathbb{R}^{C\times C}$ is averaged across [layer,head]-dimensions.}
  \State $\mathcal{P} \gets \mathcal{P} + \hat{\mathcal{P}}/N_{\mathrm{samples}}$
  \EndWhile
  \end{algorithmic}
\end{algorithm}

\subsection{BrainStratify-Fine}
\begin{figure}[h]
  \centering
  \includegraphics[width=\linewidth]{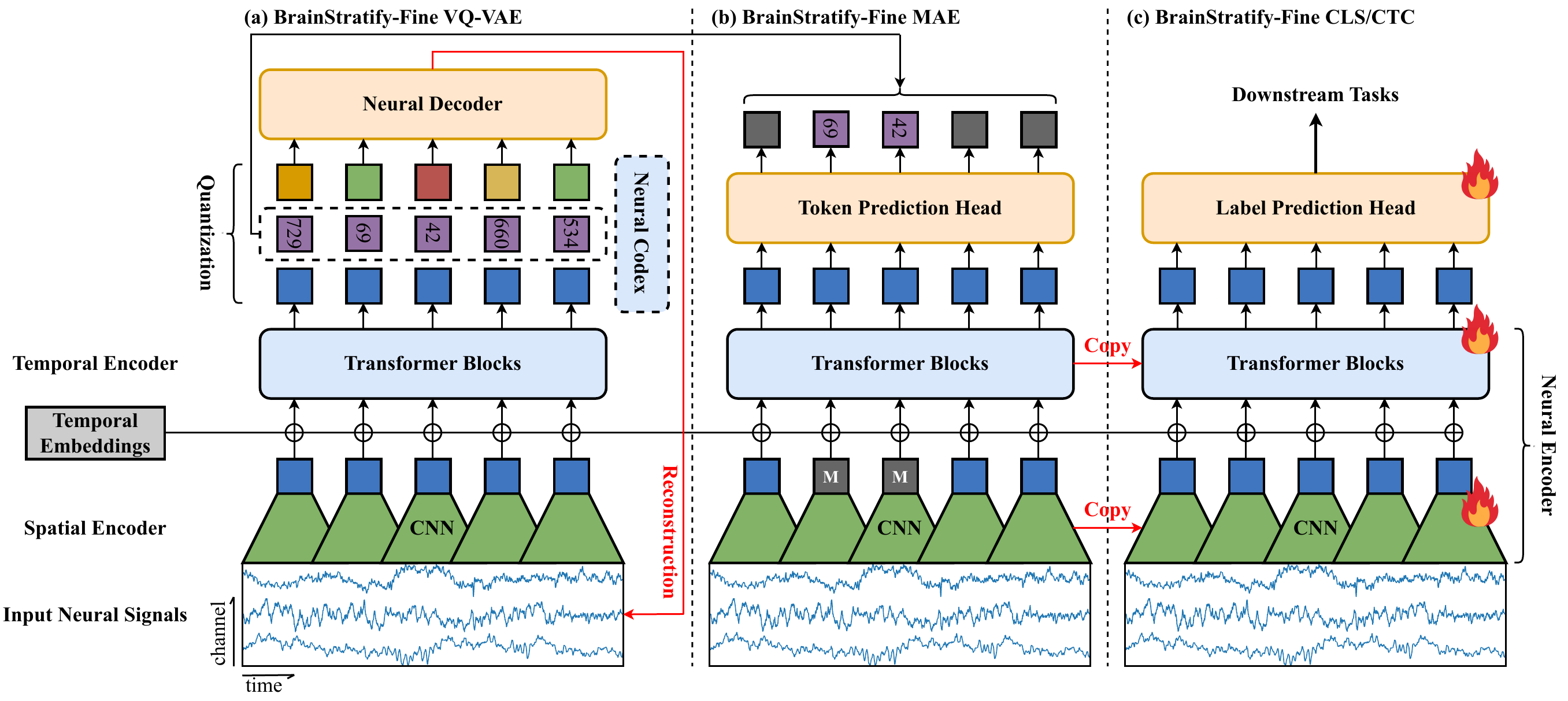}
  \caption{\textbf{An overview of the three-stage training pipeline of BrainStratify-Fine.} \textbf{(a).} Learning discrete neural codex in the BrainStratify-Fine VQ-VAE by reconstructing the original neural signals. \textbf{(b).} Mask modeling pre-training of Neural Encoder in the BrainStratify-Fine MAE. \textbf{(c).} Fine-tuning the pre-trained Neural Encoder with an MLP head for various downstream decoding tasks.}
  \label{fig:bsf}
\end{figure}

After identifying coarse-grained functional groups, BrainStratify-Fine aims to further identify fine-grained functional modules through decoupled product quantization (DPQ). The three-stage training framework of BrainStratify-Fine is illustrated in Figure \ref{fig:bsf}. "Neural Encoder" is shared across BrainStratify-Fine variants. 
The hyperparameters of Neural Encoder are shown in Table \ref{table:model-neural-encoder}.
\begin{table}[h]
  \caption{The hyperparameters of Neural Encoder.}
  \label{table:model-neural-encoder}
  \centering
  \begin{tabular}{ccc}
    \toprule
    \textbf{Module} & \textbf{Name} & \textbf{Value} \\
    \midrule
    \multirow{6}{*}{Patch Tokenizer} & Linear Projection & $C\rightarrow C$ \\
    & \# of Input Channels & \{$C$,256,256,256\} \\
    & \# of Output Channels & \{256,256,256,256\} \\
    & Kernel Size & \{9,3,3,3\} \\
    & Stride & \{5,2,2,2\} \\
    & Padding & \{4,1,1,1\} \\
    \midrule
    \multirow{7}{*}{Temporal Transformer} & \# of Transformer Layers & 8 \\
    & Hidden Size & 256 \\
    & MLP Size & 1024 \\
    & MLP Dropout Ratio & \{0.2,0.\} \\
    & \# of Attention Heads & 8 \\
    & Attention Head Size & 64 \\
    & Attention Dropout Ratio & 0.2 \\
    \bottomrule
  \end{tabular}
\end{table}
\subsubsection{BrainStratify-Fine VQ-VAE}
The architecture of BrainStratify-Fine VQ-VAE contains three parts: (1) Neural Encoder, (2) Vector Quantizer, and (3) Neural Decoder. The architecture of Neural Encoder is shown in Table \ref{table:model-neural-encoder}. The "Vector Quantizer" is modified from Du-IN \cite{zheng2025discrete} to combine product quantization \cite{jegou2010product} with partial-correlation constraint \cite{hazarika2020misa,li2025revisit}. The "Neural Decoder" contains:
\begin{itemize}
  \item \textbf{Temporal Transformer:} A stack of Transformer layers.
  \item \textbf{Time Regression (RGS) Head:} A stack of 1D Transposed Convolution layers and one linear projection layer.
\end{itemize}
The hyperparameters for BrainStratify-Fine VQ-VAE training are shown in Table \ref{table:model-brainstratify-fine-vqvae}.

\subsubsection{BrainStratify-Fine MAE}
The architecture of BrainStratify-Fine MAE contains two parts: (1) Neural Encoder, and (2) Token Classification (CLS) Head. The architecture of Neural Encoder is shown in Table \ref{table:model-neural-encoder}. It's worth noting that when training BrainStratify-Fine MAE, the weights of "Neural Encoder" are randomly initialized, instead of loaded from the pre-trained BrainStratify-Fine VQ-VAE model. The hyperparameters for BrainStratify-Fine MAE training are shown in Table \ref{table:model-brainstratify-fine-mae}.

\subsubsection{BrainStratify-Fine CLS}
The BrainStratify-Fine CLS model is designed for the classification (CLS) task -- decode the corresponding label $\bm{y}$ from a sequence of raw neural signals $\mathcal{X}$. The architecture of BrainStratify-Fine CLS contains two parts: (1) Neural Encoder, and (2) Label Classification (CLS) Head. The architecture of Neural Encoder is shown in Table \ref{table:model-neural-encoder}. It's worth noting that the weights of "Neural Encoder" in BrainStratify-Fine CLS can be loaded from either the pre-trained BrainStratify-Fine MAE or the pre-trained BrainStratify-Fine VQ-VAE. In the ablation experiments shown in Figure \ref{fig:ablation-codex}, our models have different suffixes:
\begin{itemize}
  \item \textbf{BrainStratify-Fine (vqvae):} The weights of "Neural Encoder" in BrainStratify-Fine CLS are loaded from the pre-trained BrainStratify-Fine VQ-VAE.
  \item \textbf{BrainStratify-Fine (mae):} The weights of "Neural Encoder" in BrainStratify-Fine CLS are loaded from the pre-trained BrainStratify-Fine MAE.
\end{itemize}
The "Label Classification (CLS) Head" is an MLP with one hidden layer, flattens the output embedding sequence from upstream, and maps this feature embedding to the final results through MLP. The hyperparameters for BrainStratify-Fine CLS training are shown in Table \ref{table:model-brainstratify-fine-cls}.
\subsubsection{BrainStratify-Fine CTC}
The BrainStratify-Fine CTC model is designed for the connectionist temporal classification (CTC) task -- decode the corresponding label sequence $\bm{y}$ from a sequence of raw neural signals $\mathcal{X}$; see Appendix \ref{sec:supp-task-details} for more details. The architecture of BrainStratify-Fine CTC contains two parts: (1) Neural Encoder, and (2) Label Classification (CLS) Head. The architecture of Neural Encoder is shown in Table \ref{table:model-neural-encoder}. The weights of "Neural Encoder" in BrainStratify-Fine CTC can be loaded from either the pre-trained BrainStratify-Fine MAE or the pre-trained BrainStratify-Fine VQ-VAE. The "Label Classification (CLS) Head" is an MLP with one hidden layer, flattens the output embeddings with a specified flatten window from upstream, and maps the transformed embedding sequence to the final results through MLP. The hyperparameters for BrainStratify-Fine CTC training are shown in Table \ref{table:model-brainstratify-fine-ctc}.
\begin{table}[h]
  \caption{The hyperparameters for BrainStratify-Fine VQ-VAE training.}
  \label{table:model-brainstratify-fine-vqvae}
  \centering
  \begin{tabular}{cccc}
    \toprule
    \textbf{Module} & \textbf{Sub-Module} & \textbf{Name} & \textbf{Value} \\
    \midrule
    \multirow{1}{*}{Neural Encoder} & - & - & - \\
    \midrule
    \multirow{4}{*}{\makecell[c]{Vector\\Quantizer}} & \multirow{4}{*}{-} & \# of Groups & 4 \\
    & & Codex Size per Group & $256\times 64$ \\
    & & Embedding-to-Codex Projection & $256\rightarrow 256(\mathrm{Tanh})\rightarrow 64$ \\
    & & Codex-to-Embedding Projection & $64\rightarrow 256$ \\
    \midrule
    \multirow{14}{*}{Neural Decoder} & \multirow{7}{*}{\makecell[c]{Temporal\\Transformer}} & \# of Transformer Layers & 4 \\
    & & Hidden Size & 256 \\
    & & MLP Size & 1024 \\
    & & MLP Dropout Ratio & \{0.2,0.\} \\
    & & \# of Attention Heads & 8 \\
    & & Attention Head Size & 64 \\
    & & Attention Dropout Ratio & 0.2 \\
    \cline{2-4}
    & \multirow{7}{*}{\makecell[c]{Time\\RGS Head}} & \# of Input Channels & \{256,256,256,256\} \\
    & & \# of Output Channels & \{256,256,256,256\} \\
    & & Kernel Size & \{3,3,3,9\} \\
    & & Stride & \{2,2,2,5\} \\
    & & Padding & - \\
    & & Output Padding & - \\
    & & Linear Projection & $256\rightarrow C$ \\
    \midrule
    \multirow{9}{*}{\makecell[c]{Optimizer}} & \multirow{9}{*}{-} & Batch Size & 64 \\
    & & Maximum Learning Rate & 3e-4 \\
    & & Minimum Learning Rate & 5e-5 \\
    & & Learning Rate Scheduler & Cosine \\
    & & Optimizer Type & AdamW \\
    & & Adam $\beta$ & $(0.9,0.99)$ \\
    & & Weight Decay & 0.01 \\
    & & Total Epochs & 400 \\
    & & Warm-up Epochs & 40 \\
    \bottomrule
  \end{tabular}
\end{table}
\begin{table}[htbp]
  \caption{The hyperparameters for BrainStratify-Fine MAE training.}
  \label{table:model-brainstratify-fine-mae}
  \centering
  \begin{tabular}{cccc}
    \toprule
    \textbf{Module} & \textbf{Sub-Module} & \textbf{Name} & \textbf{Value} \\
    \midrule
    \multirow{1}{*}{Neural Encoder} & - & - & - \\
    \midrule
    \multirow{2}{*}{\makecell[c]{Token CLS Head}} & \multirow{2}{*}{-} & \# of Classification Heads & 4 \\
    & & Linear Projection & $256\rightarrow 256$ \\
    \midrule
    \multirow{9}{*}{\makecell[c]{Optimizer}} & \multirow{9}{*}{-} & Batch Size & 64 \\
    & & Maximum Learning Rate & 3e-4 \\
    & & Minimum Learning Rate & 5e-5 \\
    & & Learning Rate Scheduler & Cosine \\
    & & Optimizer Type & AdamW \\
    & & Adam $\beta$ & $(0.9,0.99)$ \\
    & & Weight Decay & 0.05 \\
    & & Total Epochs & 400 \\
    & & Warm-up Epochs & 40 \\
    \bottomrule
  \end{tabular}
\end{table}
\begin{table}[htbp]
  \caption{The hyperparameters for BrainStratify-Fine CLS training.}
  \label{table:model-brainstratify-fine-cls}
  \centering
  \tabcolsep=0.01\linewidth
  \begin{tabular}{cccc}
    \toprule
    \textbf{Module} & \textbf{Sub-Module} & \textbf{Name} & \textbf{Value} \\
    \midrule
    \multirow{1}{*}{Neural Encoder} & - & - & - \\
    \midrule
    \multirow{2}{*}{\makecell[c]{Label CLS Head}} & \multirow{2}{*}{-} & Flatten & - \\
    & & Linear Projection & $N_{f}\times 256 \rightarrow 128 (\mathrm{ReLU})\rightarrow |\mathcal{Y}|$ \\
    \midrule
    \multirow{9}{*}{\makecell[c]{Optimizer}} & \multirow{9}{*}{-} & Batch Size & 32 \\
    & & Maximum Learning Rate & 2e-4 \\
    & & Minimum Learning Rate & 5e-6 \\
    & & Learning Rate Scheduler & Cosine \\
    & & Optimizer Type & AdamW \\
    & & Adam $\beta$ & $(0.9,0.99)$ \\
    & & Weight Decay & 0.05 \\
    & & Total Epochs & 200 \\
    & & Warm-up Epochs & 20 \\
    \bottomrule
  \end{tabular}
\end{table}
\begin{table}[htbp]
  \caption{The hyperparameters for BrainStratify-Fine CTC training.}
  \label{table:model-brainstratify-fine-ctc}
  \centering
  \begin{tabular}{cccc}
    \toprule
    \textbf{Module} & \textbf{Sub-Module} & \textbf{Name} & \textbf{Value} \\
    \midrule
    \multirow{1}{*}{Neural Encoder} & - & - & - \\
    \midrule
    \multirow{2}{*}{\makecell[c]{Label CLS Head}} & \multirow{2}{*}{-} & Flatten Window & 3 \\
    & & Linear Projection & $3\times 256 \rightarrow 128 (\mathrm{ReLU})\rightarrow |\mathcal{Y}|$ \\
    \midrule
    \multirow{9}{*}{\makecell[c]{Optimizer}} & \multirow{9}{*}{-} & Batch Size & 32 \\
    & & Maximum Learning Rate & 2e-4 \\
    & & Minimum Learning Rate & 5e-6 \\
    & & Learning Rate Scheduler & Cosine \\
    & & Optimizer Type & AdamW \\
    & & Adam $\beta$ & $(0.9,0.99)$ \\
    & & Weight Decay & 0.05 \\
    & & Total Epochs & 200 \\
    & & Warm-up Epochs & 20 \\
    \bottomrule
  \end{tabular}
\end{table}

\clearpage
\section{Data Augmentation}
\label{sec:supp-data-augmentation}
To enhance the robustness of learned representations during both the pre-training and fine-tuning stages, we apply data augmentation in both datasets.
\paragraph{Pre-training Dataset.}In our implementation, we segment neural recordings into 8-second samples with a 4-second overlap. When fetching a sample, we randomly select a starting point between 0 and 4 seconds, then extract a 4-second sample beginning from that point.

\paragraph{Downstream Dataset.}Since trials occur consecutively without gaps, employing the jittering mentioned above leads to the blending of information from other trials. In our implementation, we segment sEEG recordings into samples with the corresponding trial length; see Appendix \ref{sec:supp-task-details} for details. When fetching a sample, we randomly choose a shift step between 0 and 0.2 seconds, then shift the sample either to the left or right, padding it with zeros.

\section{Visualization of Spatial Context Classification}
\begin{figure}[h]
  \centering
  \includegraphics[width=\linewidth]{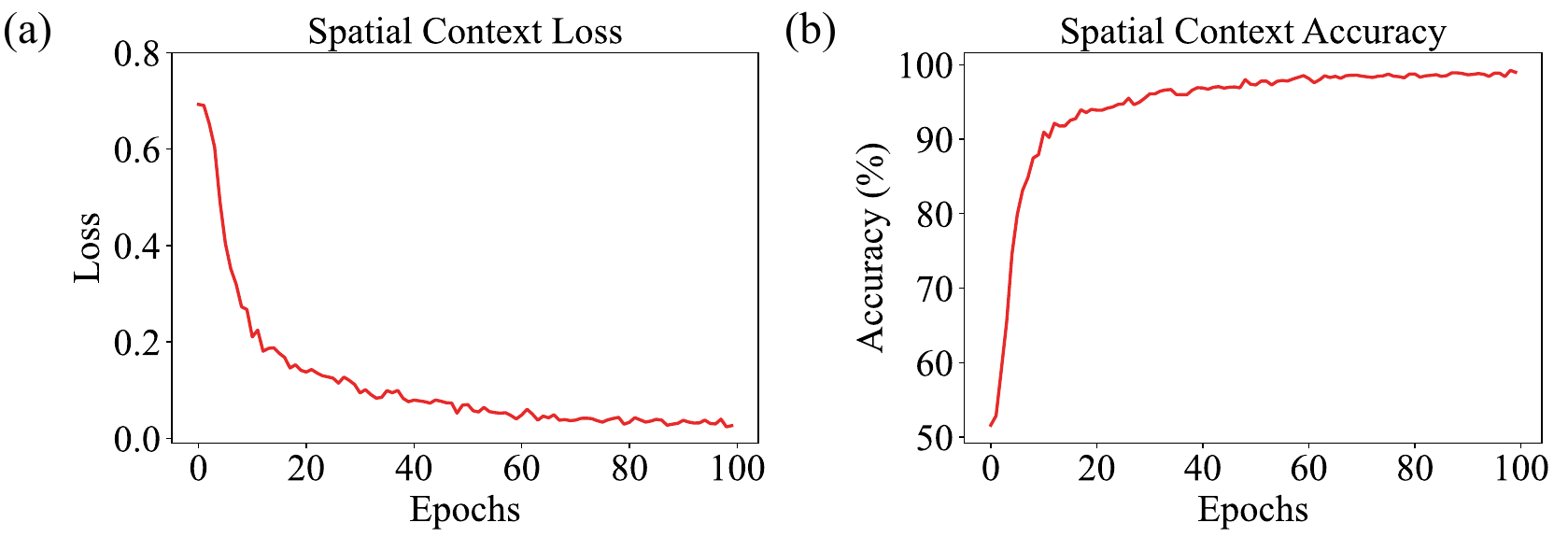}
  \caption{The loss curve and accuracy curve during the training process of BrainStratify-Coarse.}
  \label{fig:ctx-visualization}
\end{figure}

Figure \ref{fig:ctx-visualization} demonstrates the convergence curves of the total pre-training loss and spatial context accuracy of BrainStratify-Coarse. We observe that there is a stable decrease in the spatial context loss, and the spatial context accuracy achieves $\geq$ 95\%.

\section{Visualization of Vector-Quantized Neural Reconstruction}
\begin{figure}[h]
  \centering
  \includegraphics[width=\linewidth]{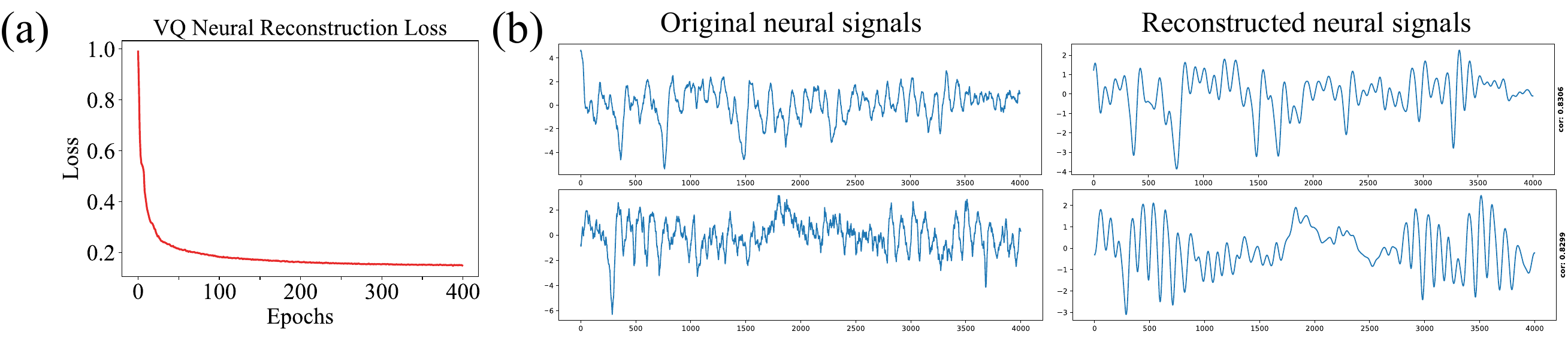}
  \caption{\textbf{The visualization of Vector-Quantized Neural Reconstruction.} \textbf{(a).} The reconstruction loss curve during the training process of BrainStratify-Fine VQ-VAE. \textbf{(b).} The visualization of reconstructed neural signals.}
  \label{fig:vqvae-visualization}
\end{figure}

We further visualize how the neural signals are reconstructed. As depicted in Figure \ref{fig:vqvae-visualization}, although some details are missing, the overall trend of the signals is reconstructed well. Meanwhile, there is a stable decrease in the reconstruction loss during training, which indicates the discrete codex does learn high-level information from neural signals.

\section{Visualization of Mask Neural Modeling}
\begin{figure}[h]
  \centering
  \includegraphics[width=\linewidth]{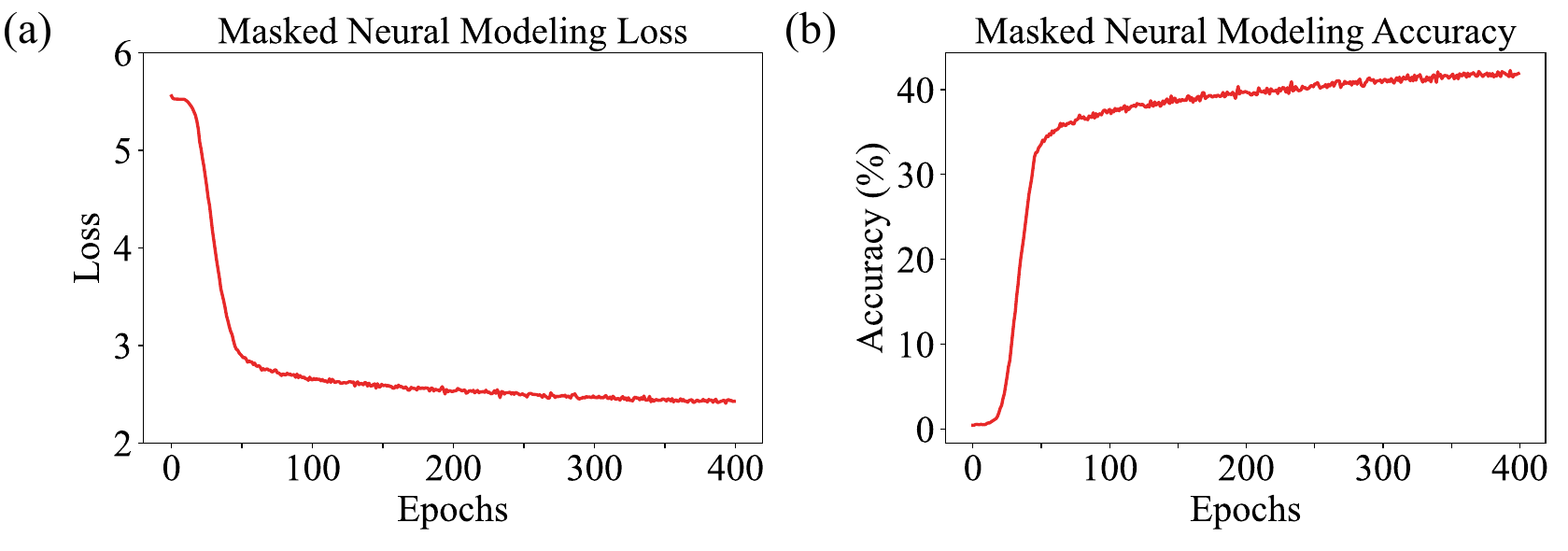}
  \caption{The loss curve and accuracy curve during the training process of BrainStratify-Fine MAE.}
  \label{fig:mae-visualization}
\end{figure}

Figure \ref{fig:mae-visualization} demonstrates the convergence curves of the total pre-training loss and masked neural modeling accuracy of BrainStratify-Fine MAE. We observe that there is a stable decrease in the mask modeling loss, and the mask modeling accuracy achieves about 40\%.

\section{Additional Ablation Study}
\label{sec:supp-ablation-study}
\begin{figure}[h]
  \centering
  \includegraphics[width=\linewidth]{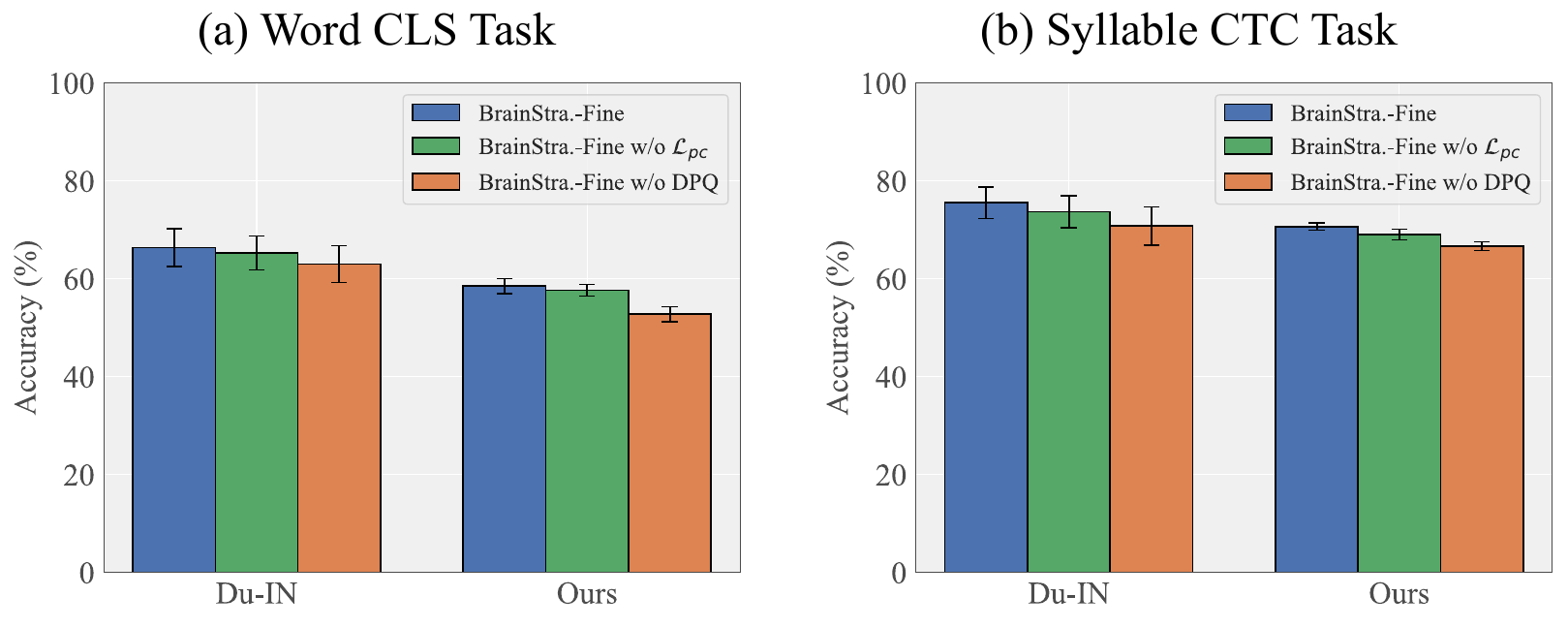}
  \caption{BrainStratify-Fine ablation study on two word-reading intracranial neural datasets (sEEG \& ECoG). The channel set selected via the Multi-Channel (MC) strategy is used for evaluation. Accuracy (\%) $\pm$ ste (\%) on Du-IN dataset \cite{zheng2025discrete} (sEEG) and our collected ECoG dataset are reported.}
  \label{fig:ablation-pcpq}
\end{figure}

In this experiment, we ablate the partial-correlation constraint ($\mathcal{L}_{pc}$) and the entire DPQ module, as illustrated in Figure \ref{fig:ablation-pcpq}. In implementation, we initialize the embedding-to-codex projection for each sub-quantizer separately, which mimics the regularization effect of $\mathcal{L}_{pc}$. Therefore, the ablation of $\mathcal{L}_{pc}$ causes only minor performance drops. However, removing DPQ entirely reduces BrainStratify-Fine to Du-IN \cite{zheng2025discrete}, significantly degrading decoding performance.

\section{Broader Impacts}
\label{supp:supp-broader-impacts}
BrainStratify has the potential to advance speech decoding and invasive BCI systems by providing a robust and interpretable solution for speech decoding from invasive recordings. Its advanced performance across datasets and modalities while keeping data-efficient makes it well-suited for real-world scenarios where large-scale medical data annotation is often prohibitively costly or unfeasible. Furthermore, our (epidural) ECoG results demonstrate the clinical viability of our method in patients with communication and functional impairments caused by amyotrophic lateral sclerosis (ALS).

\clearpage
\section{Subject-Wise Evaluation}
\label{sec:supp-subject-evaluation}
\subsection{Channel Selection Strategy}
\begin{table}[h]
  \caption{The 61-word performance on Du-IN dataset from subjects (01-04).}
  \label{table:result-ccm-subjs-1}
  \centering
  \begin{tabular}{l|c|cccc}
    \toprule
    \multirow{2}{*}{\textbf{Channel Selection}} & \multirow{2}{*}{\textbf{Supervised}} & \multicolumn{4}{c}{\textbf{Accuracy (\%) $\pm$ std (\%)}} \\
    & & subj-01 & subj-02 & subj-03 & subj-04 \\
    \midrule
    SC \cite{wang2023brainbert}    & \Checkmark   & 66.22$\pm$0.83 & 76.21$\pm$0.56 & 20.02$\pm$1.70 & 58.36$\pm$2.32 \\
    MC \cite{zheng2025discrete}    & \Checkmark   & \textbf{72.76$\pm$1.72} & \underline{78.95$\pm$0.72} & \underline{35.31$\pm$1.05} & \textbf{66.04$\pm$1.75} \\
    \midrule
    CCM \cite{chen2025similarity}  & \XSolidBrush & 50.89$\pm$2.65 & 65.32$\pm$1.36 & 7.53$\pm$0.94 & 23.44$\pm$4.66 \\
    DUET \cite{qiu2024duet}        & \XSolidBrush & 61.54$\pm$1.90 & 69.42$\pm$0.80 & 32.73$\pm$1.99 & 48.71$\pm$2.43 \\
    PopT \cite{chau2024population} & \XSolidBrush & 70.08$\pm$1.77 & 78.61$\pm$1.98 & 28.48$\pm$2.77 & 63.72$\pm$1.97 \\
    \midrule
    BrainStra.-Coarse              & \XSolidBrush & \underline{70.82$\pm$1.01} & \textbf{78.97$\pm$0.67} & \textbf{36.73$\pm$0.90} & \underline{64.75$\pm$2.24} \\
    \bottomrule
  \end{tabular}
\end{table}

\begin{table}[h]
  \caption{The 61-word performance on Du-IN dataset from subjects (05-08).}
  \label{table:result-ccm-subjs-2}
  \centering
  \begin{tabular}{l|c|cccc}
    \toprule
    \multirow{2}{*}{\textbf{Channel Selection}} & \multirow{2}{*}{\textbf{Supervised}} & \multicolumn{4}{c}{\textbf{Accuracy (\%) $\pm$ std (\%)}} \\
    & & subj-05 & subj-06 & subj-07 & subj-08 \\
    \midrule
    SC \cite{wang2023brainbert}    & \Checkmark   & 67.16$\pm$0.44 & 26.04$\pm$1.39 & 37.40$\pm$1.44 & 42.21$\pm$1.01 \\
    MC \cite{zheng2025discrete}    & \Checkmark   & \textbf{77.69$\pm$1.27} & \textbf{42.83$\pm$1.30} & \textbf{59.63$\pm$2.20} & \textbf{49.18$\pm$0.84} \\
    \midrule
    CCM \cite{chen2025similarity}  & \XSolidBrush & 46.26$\pm$3.04 & 11.55$\pm$1.74 & 28.14$\pm$2.31 & 21.88$\pm$4.26 \\
    DUET \cite{qiu2024duet}        & \XSolidBrush & 57.62$\pm$1.55 & 33.07$\pm$1.96 & 33.68$\pm$1.66 & 32.92$\pm$1.16 \\
    PopT \cite{chau2024population} & \XSolidBrush & 51.07$\pm$1.31 & 39.17$\pm$1.29 & 46.88$\pm$1.77 & 47.22$\pm$2.49 \\
    \midrule
    BrainStra.-Coarse              & \XSolidBrush & \underline{76.39$\pm$0.56} & \underline{41.64$\pm$1.73} & \underline{58.08$\pm$1.77} & \underline{47.65$\pm$1.28} \\
    \bottomrule
  \end{tabular}
\end{table}

\begin{table}[h]
  \caption{The 61-word performance on Du-IN dataset from subjects (09-12).}
  \label{table:result-ccm-subjs-3}
  \centering
  \begin{tabular}{l|c|cccc}
    \toprule
    \multirow{2}{*}{\textbf{Channel Selection}} & \multirow{2}{*}{\textbf{Supervised}} & \multicolumn{4}{c}{\textbf{Accuracy (\%) $\pm$ std (\%)}} \\
    & & subj-09 & subj-10 & subj-11 & subj-12 \\
    \midrule
    SC \cite{wang2023brainbert}    & \Checkmark   & 60.14$\pm$1.62 & 14.70$\pm$0.69 & 59.16$\pm$1.38 & 48.95$\pm$3.96 \\
    MC \cite{zheng2025discrete}    & \Checkmark   & \textbf{60.79$\pm$1.91} & 30.59$\pm$0.73 & \underline{67.12$\pm$1.44} & 58.01$\pm$3.29 \\
    \midrule
    CCM \cite{chen2025similarity}  & \XSolidBrush & 36.68$\pm$3.14 & 7.62$\pm$0.98 & 39.57$\pm$3.20 & 14.16$\pm$2.50 \\
    DUET \cite{qiu2024duet}        & \XSolidBrush & 58.44$\pm$1.07 & 30.06$\pm$1.79 & 45.06$\pm$2.20 & 38.48$\pm$3.16 \\
    PopT \cite{chau2024population} & \XSolidBrush & 58.44$\pm$1.07 & \underline{32.46$\pm$1.27} & \textbf{67.71$\pm$1.80} & \underline{58.29$\pm$1.97} \\
    \midrule
    BrainStra.-Coarse              & \XSolidBrush & \underline{60.60$\pm$0.69} & \textbf{32.60$\pm$1.27} & 67.02$\pm$1.84 & \textbf{58.29$\pm$1.97} \\
    \bottomrule
  \end{tabular}
\end{table}

\clearpage
\subsection{Results on Brain Treebank.}
\begin{table}[h]
  \caption{The performance on Brain Treebank dataset from subject 01.}
  \label{table:result-braintree-subj-1}
  \centering
  \footnotesize
  \tabcolsep=0.005\linewidth
  \begin{tabular}{l|c|cccc}
    \toprule
    \multirow{2}{*}{\textbf{Methods}} & \multirow{2}{*}{\textbf{Chan. Select.}} & \multicolumn{4}{c}{\textbf{ROC-AUC $\pm$ ste}} \\
    & & Pitch & Volumn & Sent. Onset & Word Onset \\
    \midrule
    LaBraM \cite{jiang2024large}     & BrainStra.-Coarse & 0.8027$\pm$0.0085 & 0.9040$\pm$0.0108 & 0.9327$\pm$0.0043 & 0.9663$\pm$0.0027 \\
    CBraMod \cite{wang2024cbramod}   & BrainStra.-Coarse & 0.7805$\pm$0.0040 & 0.8865$\pm$0.0062 & 0.9287$\pm$0.0073 & 0.9703$\pm$0.0009 \\
    PopT \cite{chau2024population}   & BrainStra.-Coarse & 0.8257$\pm$0.0131 & 0.9192$\pm$0.0060 & 0.9488$\pm$0.0081 & 0.9698$\pm$0.0034 \\
    \midrule
    EEG-CFMR \cite{song2022eeg}      & BrainStra.-Coarse & 0.8525$\pm$0.0039 & 0.9289$\pm$0.0089 & 0.9520$\pm$0.0049 & 0.9746$\pm$0.0012 \\
    Du-IN \cite{zheng2025discrete}   & BrainStra.-Coarse & \underline{0.8724$\pm$0.0098} & \underline{0.9414$\pm$0.0056} & \underline{0.9642$\pm$0.0031} & \underline{0.9758$\pm$0.0024} \\
    H2DiLR \cite{wu2024towards}      & BrainStra.-Coarse & 0.8450$\pm$0.0068 & 0.9225$\pm$0.0072 & 0.8943$\pm$0.0072 & 0.9170$\pm$0.0052 \\
    \midrule
    BrainStra.-Fine                  & BrainStra.-Coarse & \textbf{0.8826$\pm$0.0084} & \textbf{0.9489$\pm$0.0047} & \textbf{0.9737$\pm$0.0035} & \textbf{0.9882$\pm$0.0018} \\
    \bottomrule
  \end{tabular}
\end{table}

\begin{table}[h]
  \caption{The performance on Brain Treebank dataset from subject 02.}
  \label{table:result-braintree-subj-2}
  \centering
  \footnotesize
  \tabcolsep=0.005\linewidth
  \begin{tabular}{l|c|cccc}
    \toprule
    \multirow{2}{*}{\textbf{Methods}} & \multirow{2}{*}{\textbf{Chan. Select.}} & \multicolumn{4}{c}{\textbf{ROC-AUC $\pm$ ste}} \\
    & & Pitch & Volumn & Sent. Onset & Word Onset \\
    \midrule
    LaBraM \cite{jiang2024large}     & BrainStra.-Coarse & 0.7852$\pm$0.0111 & 0.9307$\pm$0.0076 & 0.7868$\pm$0.0156 & 0.9227$\pm$0.0054 \\
    CBraMod \cite{wang2024cbramod}   & BrainStra.-Coarse & 0.7861$\pm$0.0164 & 0.9351$\pm$0.0038 & 0.7836$\pm$0.0136 & 0.9306$\pm$0.0027 \\
    PopT \cite{chau2024population}   & BrainStra.-Coarse & 0.7748$\pm$0.0203 & 0.9512$\pm$0.0122 & 0.8579$\pm$0.0174 & 0.9436$\pm$0.0053 \\
    \midrule
    EEG-CFMR \cite{song2022eeg}      & BrainStra.-Coarse & 0.8164$\pm$0.0142 & 0.9530$\pm$0.0089 & 0.8304$\pm$0.0139 & 0.9646$\pm$0.0026 \\
    Du-IN \cite{zheng2025discrete}   & BrainStra.-Coarse & 0.8130$\pm$0.0188 & \underline{0.9586$\pm$0.0085} & \textbf{0.8752$\pm$0.0097} & \underline{0.9694$\pm$0.0018} \\
    H2DiLR \cite{wu2024towards}      & BrainStra.-Coarse & \underline{0.8225$\pm$0.0077} & 0.9433$\pm$0.0105 & 0.7860$\pm$0.0212 & 0.9016$\pm$0.0047 \\
    \midrule
    BrainStra.-Fine                  & BrainStra.-Coarse & \textbf{0.8292$\pm$0.0125} & \textbf{0.9705$\pm$0.0062} & \underline{0.8582$\pm$0.0095} & \textbf{0.9806$\pm$0.0041} \\
    \bottomrule
  \end{tabular}
\end{table}

\begin{table}[h]
  \caption{The performance on Brain Treebank dataset from subject 03.}
  \label{table:result-braintree-subj-3}
  \centering
  \footnotesize
  \tabcolsep=0.005\linewidth
  \begin{tabular}{l|c|cccc}
    \toprule
    \multirow{2}{*}{\textbf{Methods}} & \multirow{2}{*}{\textbf{Chan. Select.}} & \multicolumn{4}{c}{\textbf{ROC-AUC $\pm$ ste}} \\
    & & Pitch & Volumn & Sent. Onset & Word Onset \\
    \midrule
    LaBraM \cite{jiang2024large}     & BrainStra.-Coarse & 0.6973$\pm$0.0141 & 0.8971$\pm$0.0060 & 0.9614$\pm$0.0044 & 0.9590$\pm$0.0076 \\
    CBraMod \cite{wang2024cbramod}   & BrainStra.-Coarse & 0.6734$\pm$0.0084 & 0.8712$\pm$0.0048 & 0.9565$\pm$0.0065 & 0.9477$\pm$0.0071 \\
    PopT \cite{chau2024population}   & BrainStra.-Coarse & 0.7301$\pm$0.0185 & 0.9078$\pm$0.0068 & 0.9648$\pm$0.0042 & 0.9605$\pm$0.0069 \\
    \midrule
    EEG-CFMR \cite{song2022eeg}      & BrainStra.-Coarse & 0.7537$\pm$0.0211 & 0.9188$\pm$0.0101 & 0.9727$\pm$0.0034 & 0.9582$\pm$0.0039 \\
    Du-IN \cite{zheng2025discrete}   & BrainStra.-Coarse & \underline{0.7593$\pm$0.0127} & \underline{0.9343$\pm$0.0041} & \underline{0.9737$\pm$0.0034} & \underline{0.9639$\pm$0.0037} \\
    H2DiLR \cite{wu2024towards}      & BrainStra.-Coarse & 0.7405$\pm$0.0239 & 0.9066$\pm$0.0041 & 0.9293$\pm$0.0078 & 0.9112$\pm$0.0064 \\
    \midrule
    BrainStra.-Fine                  & BrainStra.-Coarse & \textbf{0.7696$\pm$0.0193} & \textbf{0.9411$\pm$0.0050} & \textbf{0.9833$\pm$0.0021} & \textbf{0.9766$\pm$0.0024} \\
    \bottomrule
  \end{tabular}
\end{table}

\begin{table}[h]
  \caption{The performance on Brain Treebank dataset from subject 04.}
  \label{table:result-braintree-subj-4}
  \centering
  \footnotesize
  \tabcolsep=0.005\linewidth
  \begin{tabular}{l|c|cccc}
    \toprule
    \multirow{2}{*}{\textbf{Methods}} & \multirow{2}{*}{\textbf{Chan. Select.}} & \multicolumn{4}{c}{\textbf{ROC-AUC $\pm$ ste}} \\
    & & Pitch & Volumn & Sent. Onset & Word Onset \\
    \midrule
    LaBraM \cite{jiang2024large}     & BrainStra.-Coarse & 0.7211$\pm$0.0200 & 0.8392$\pm$0.0115 & 0.8666$\pm$0.0131 & 0.9081$\pm$0.0080 \\
    CBraMod \cite{wang2024cbramod}   & BrainStra.-Coarse & 0.6969$\pm$0.0167 & 0.7888$\pm$0.0077 & 0.8582$\pm$0.0145 & 0.8969$\pm$0.0083 \\
    PopT \cite{chau2024population}   & BrainStra.-Coarse & 0.7542$\pm$0.0158 & 0.8484$\pm$0.0219 & \textbf{0.9326$\pm$0.0058} & 0.9351$\pm$0.0040 \\
    \midrule
    EEG-CFMR \cite{song2022eeg}      & BrainStra.-Coarse & 0.7546$\pm$0.0236 & 0.8394$\pm$0.0135 & 0.8825$\pm$0.0132 & 0.8631$\pm$0.0076 \\
    Du-IN \cite{zheng2025discrete}   & BrainStra.-Coarse & \underline{0.7619$\pm$0.0098} & \underline{0.8707$\pm$0.0132} & \underline{0.9317$\pm$0.0093} & \underline{0.9613$\pm$0.0026} \\
    H2DiLR \cite{wu2024towards}      & BrainStra.-Coarse & 0.7585$\pm$0.0081 & 0.8418$\pm$0.0085 & 0.8108$\pm$0.0149 & 0.8086$\pm$0.0118 \\
    \midrule
    BrainStra.-Fine                  & BrainStra.-Coarse & \textbf{0.7656$\pm$0.0156} & \textbf{0.8762$\pm$0.0168} & 0.9176$\pm$0.0068 & \textbf{0.9621$\pm$0.0041} \\
    \bottomrule
  \end{tabular}
\end{table}

\begin{table}[h]
  \caption{The performance on Brain Treebank dataset from subject 05.}
  \label{table:result-braintree-subj-5}
  \centering
  \footnotesize
  \tabcolsep=0.005\linewidth
  \begin{tabular}{l|c|cccc}
    \toprule
    \multirow{2}{*}{\textbf{Methods}} & \multirow{2}{*}{\textbf{Chan. Select.}} & \multicolumn{4}{c}{\textbf{ROC-AUC $\pm$ ste}} \\
    & & Pitch & Volumn & Sent. Onset & Word Onset \\
    \midrule
    LaBraM \cite{jiang2024large}     & BrainStra.-Coarse & 0.6962$\pm$0.0144 & 0.8601$\pm$0.0067 & 0.9561$\pm$0.0064 & 0.9651$\pm$0.0027 \\
    CBraMod \cite{wang2024cbramod}   & BrainStra.-Coarse & 0.6766$\pm$0.0084 & 0.8298$\pm$0.0129 & 0.9525$\pm$0.0088 & 0.9590$\pm$0.0053 \\
    PopT \cite{chau2024population}   & BrainStra.-Coarse & 0.6992$\pm$0.0102 & 0.8722$\pm$0.0157 & \underline{0.9776$\pm$0.0052} & 0.9771$\pm$0.0032 \\
    \midrule
    EEG-CFMR \cite{song2022eeg}      & BrainStra.-Coarse & 0.7297$\pm$0.0181 & 0.8719$\pm$0.0093 & 0.9632$\pm$0.0054 & 0.9695$\pm$0.0042 \\
    Du-IN \cite{zheng2025discrete}   & BrainStra.-Coarse & \underline{0.7397$\pm$0.0188} & \underline{0.8822$\pm$0.0077} & 0.9739$\pm$0.0029 & \underline{0.9784$\pm$0.0025} \\
    H2DiLR \cite{wu2024towards}      & BrainStra.-Coarse & 0.7244$\pm$0.0078 & 0.8571$\pm$0.0092 & 0.9136$\pm$0.0074 & 0.9126$\pm$0.0069 \\
    \midrule
    BrainStra.-Fine                  & BrainStra.-Coarse & \textbf{0.7416$\pm$0.0189} & \textbf{0.8862$\pm$0.0089} & \textbf{0.9849$\pm$0.0043} & \textbf{0.9907$\pm$0.0023} \\
    \bottomrule
  \end{tabular}
\end{table}

\begin{table}[h]
  \caption{The performance on Brain Treebank dataset from subject 06.}
  \label{table:result-braintree-subj-6}
  \centering
  \footnotesize
  \tabcolsep=0.005\linewidth
  \begin{tabular}{l|c|cccc}
    \toprule
    \multirow{2}{*}{\textbf{Methods}} & \multirow{2}{*}{\textbf{Chan. Select.}} & \multicolumn{4}{c}{\textbf{ROC-AUC $\pm$ ste}} \\
    & & Pitch & Volumn & Sent. Onset & Word Onset \\
    \midrule
    LaBraM \cite{jiang2024large}     & BrainStra.-Coarse & 0.6574$\pm$0.0093 & 0.7326$\pm$0.0285 & 0.8799$\pm$0.0073 & 0.9352$\pm$0.0074 \\
    CBraMod \cite{wang2024cbramod}   & BrainStra.-Coarse & 0.6088$\pm$0.0152 & 0.7041$\pm$0.0226 & 0.8599$\pm$0.0100 & 0.9180$\pm$0.0056 \\
    PopT \cite{chau2024population}   & BrainStra.-Coarse & 0.6929$\pm$0.0279 & 0.7610$\pm$0.0454 & \underline{0.9207$\pm$0.0205} & 0.9498$\pm$0.0090 \\
    \midrule
    EEG-CFMR \cite{song2022eeg}      & BrainStra.-Coarse & 0.7398$\pm$0.0233 & 0.7998$\pm$0.0212 & 0.8834$\pm$0.0098 & 0.9391$\pm$0.0063 \\
    Du-IN \cite{zheng2025discrete}   & BrainStra.-Coarse & \underline{0.7728$\pm$0.0255} & \underline{0.8152$\pm$0.0225} & 0.9197$\pm$0.0179 & \underline{0.9615$\pm$0.0063} \\
    H2DiLR \cite{wu2024towards}      & BrainStra.-Coarse & 0.7414$\pm$0.0255 & 0.7859$\pm$0.0302 & 0.8636$\pm$0.0168 & 0.9035$\pm$0.0100 \\
    \midrule
    BrainStra.-Fine                  & BrainStra.-Coarse & \textbf{0.7734$\pm$0.0114} & \textbf{0.8303$\pm$0.0271} & \textbf{0.9296$\pm$0.0044} & \textbf{0.9811$\pm$0.0052} \\
    \bottomrule
  \end{tabular}
\end{table}

\begin{table}[h]
  \caption{The performance on Brain Treebank dataset from subject 10.}
  \label{table:result-braintree-subj-10}
  \centering
  \footnotesize
  \tabcolsep=0.005\linewidth
  \begin{tabular}{l|c|cccc}
    \toprule
    \multirow{2}{*}{\textbf{Methods}} & \multirow{2}{*}{\textbf{Chan. Select.}} & \multicolumn{4}{c}{\textbf{ROC-AUC $\pm$ ste}} \\
    & & Pitch & Volumn & Sent. Onset & Word Onset \\
    \midrule
    LaBraM \cite{jiang2024large}     & BrainStra.-Coarse & 0.7098$\pm$0.0101 & 0.8919$\pm$0.0063 & 0.9320$\pm$0.0085 & 0.9270$\pm$0.0073 \\
    CBraMod \cite{wang2024cbramod}   & BrainStra.-Coarse & 0.6889$\pm$0.0121 & 0.8650$\pm$0.0068 & 0.9285$\pm$0.0065 & 0.9120$\pm$0.0080 \\
    PopT \cite{chau2024population}   & BrainStra.-Coarse & 0.7296$\pm$0.0330 & 0.9063$\pm$0.0052 & 0.9506$\pm$0.0057 & 0.9497$\pm$0.0029 \\
    \midrule
    EEG-CFMR \cite{song2022eeg}      & BrainStra.-Coarse & 0.7573$\pm$0.0172 & 0.9170$\pm$0.0061 & 0.9324$\pm$0.0112 & 0.9323$\pm$0.0072 \\
    Du-IN \cite{zheng2025discrete}   & BrainStra.-Coarse & \underline{0.7781$\pm$0.0194} & \underline{0.9340$\pm$0.0086} & \underline{0.9619$\pm$0.0056} & \underline{0.9580$\pm$0.0045} \\
    H2DiLR \cite{wu2024towards}      & BrainStra.-Coarse & 0.7633$\pm$0.0229 & 0.9095$\pm$0.0121 & 0.9469$\pm$0.0121 & 0.9252$\pm$0.0077 \\
    \midrule
    BrainStra.-Fine                  & BrainStra.-Coarse & \textbf{0.7837$\pm$0.0145} & \textbf{0.9380$\pm$0.0054} & \textbf{0.9725$\pm$0.0059} & \textbf{0.9744$\pm$0.0043} \\
    \bottomrule
  \end{tabular}
\end{table}

\clearpage
\subsection{Results on Du-IN dataset.}
\begin{table}[h]
  \caption{The 61-word performance on Du-IN dataset from subjects (01-04).}
  \label{table:result-duin-word-subjs-1}
  \centering
  \tabcolsep=0.01\linewidth
  \begin{tabular}{l|c|cccc}
    \toprule
    \multirow{2}{*}{\textbf{Methods}} & \multirow{2}{*}{\textbf{Chan. Select.}} & \multicolumn{4}{c}{\textbf{Accuracy (\%) $\pm$ std (\%)}} \\
    & & subj-01 & subj-02 & subj-03 & subj-04 \\
    \midrule
    LaBraM \cite{jiang2024large}   & MC & 14.14$\pm$1.28 & 39.50$\pm$1.35 & 3.28$\pm$0.55 & 12.77$\pm$1.72 \\
    CBraMod \cite{wang2024cbramod} & MC & 13.83$\pm$2.20 & 37.28$\pm$4.78 & 3.67$\pm$0.57 & 11.86$\pm$0.94 \\
    PopT \cite{chau2024population} & MC & 28.15$\pm$1.77 & 44.00$\pm$2.27 & 5.65$\pm$0.80 & 37.71$\pm$2.58 \\
    \midrule
    EEG-CFMR \cite{song2022eeg}    & MC & 58.41$\pm$1.03 & 69.82$\pm$1.22 & 19.50$\pm$1.71 & 49.65$\pm$2.38 \\
    Du-IN \cite{zheng2025discrete} & MC & \underline{78.60$\pm$0.79} & \textbf{83.61$\pm$0.38} & 38.80$\pm$2.52 & 70.98$\pm$0.81 \\
    H2DiLR \cite{wu2024towards}    & MC & 36.30$\pm$0.67 & 48.04$\pm$1.56 & 10.51$\pm$1.63 & 28.63$\pm$1.10 \\
    \midrule
    BrainStra.-Fine                & MC                & \textbf{79.66$\pm$1.61} & \underline{83.29$\pm$0.65} & \underline{46.40$\pm$3.15} & \textbf{75.78$\pm$0.94} \\
    BrainStra.-Fine                & BrainStra.-Coarse & 76.38$\pm$1.13 & 82.72$\pm$1.80 & \textbf{52.39$\pm$2.18} & \underline{75.64$\pm$1.95} \\
    \bottomrule
  \end{tabular}
\end{table}

\begin{table}[h]
  \caption{The 61-word performance on Du-IN dataset from subjects (05-08).}
  \label{table:result-duin-word-subjs-2}
  \centering
  \tabcolsep=0.01\linewidth
  \begin{tabular}{l|c|cccc}
    \toprule
    \multirow{2}{*}{\textbf{Methods}} & \multirow{2}{*}{\textbf{Chan. Select.}} & \multicolumn{4}{c}{\textbf{Accuracy (\%) $\pm$ std (\%)}} \\
    & & subj-05 & subj-06 & subj-07 & subj-08 \\
    \midrule
    LaBraM \cite{jiang2024large}   & MC & 17.73$\pm$1.11 & 5.81$\pm$1.61 & 12.90$\pm$1.26 & 6.41$\pm$1.85 \\
    CBraMod \cite{wang2024cbramod} & MC & 16.09$\pm$1.13 & 7.29$\pm$0.59 & 8.74$\pm$0.88 & 7.74$\pm$0.58 \\
    PopT \cite{chau2024population} & MC & 19.02$\pm$1.42 & 17.65$\pm$0.97 & 22.94$\pm$1.45 & 20.54$\pm$1.18 \\
    \midrule
    EEG-CFMR \cite{song2022eeg}    & MC & 65.44$\pm$1.31 & 31.06$\pm$2.58 & 47.89$\pm$1.86 & 42.12$\pm$2.08 \\
    Du-IN \cite{zheng2025discrete} & MC & \textbf{81.56$\pm$1.11} & 46.90$\pm$1.02 & \underline{65.45$\pm$1.74} & \textbf{59.09$\pm$0.98} \\
    H2DiLR \cite{wu2024towards}    & MC & 39.92$\pm$1.22 & 17.79$\pm$1.07 & 23.36$\pm$0.48 & 22.13$\pm$1.14 \\
    \midrule
    BrainStra.-Fine                & MC                & \underline{81.37$\pm$1.61} & \textbf{49.03$\pm$0.79} & \textbf{66.31$\pm$0.67} & \underline{58.99$\pm$1.26} \\
    BrainStra.-Fine                & BrainStra.-Coarse & 80.66$\pm$1.61 & \underline{49.01$\pm$2.04} & 61.87$\pm$1.44 & 58.84$\pm$0.94 \\
    \bottomrule
  \end{tabular}
\end{table}

\begin{table}[h]
  \caption{The 61-word performance on Du-IN dataset from subjects (09-12).}
  \label{table:result-duin-word-subjs-3}
  \centering
  \tabcolsep=0.01\linewidth
  \begin{tabular}{l|c|cccc}
    \toprule
    \multirow{2}{*}{\textbf{Methods}} & \multirow{2}{*}{\textbf{Chan. Select.}} & \multicolumn{4}{c}{\textbf{Accuracy (\%) $\pm$ std (\%)}} \\
    & & subj-09 & subj-10 & subj-11 & subj-12 \\
    \midrule
    LaBraM \cite{jiang2024large}   & MC & 9.35$\pm$1.09 & 3.91$\pm$0.31 & 7.84$\pm$0.92 & 7.74$\pm$1.24 \\
    CBraMod \cite{wang2024cbramod} & MC & 13.70$\pm$2.33 & 5.62$\pm$0.40 & 6.04$\pm$0.65 & 6.38$\pm$1.41 \\
    PopT \cite{chau2024population} & MC & 32.43$\pm$2.64 & 5.07$\pm$2.48 & 23.53$\pm$2.99 & 13.93$\pm$0.95 \\
    \midrule
    EEG-CFMR \cite{song2022eeg}    & MC & 56.51$\pm$1.98 & 22.22$\pm$1.07 & 57.10$\pm$2.03 & 29.87$\pm$1.44 \\
    Du-IN \cite{zheng2025discrete} & MC & 69.18$\pm$1.96 & 34.23$\pm$1.17 & 75.52$\pm$1.27 & 48.54$\pm$0.56 \\
    H2DiLR \cite{wu2024towards}    & MC & 25.99$\pm$0.52 & 10.68$\pm$0.98 & 27.07$\pm$1.61 & 19.63$\pm$1.14 \\
    \midrule
    BrainStra.-Fine                & MC                & \underline{74.78$\pm$0.36} & \underline{39.17$\pm$1.81} & \underline{77.85$\pm$0.71} & \textbf{63.61$\pm$1.64} \\
    BrainStra.-Fine                & BrainStra.-Coarse & \textbf{77.43$\pm$1.17} & \textbf{39.98$\pm$2.18} & \textbf{78.92$\pm$0.99} & \underline{63.43$\pm$2.02} \\
    \bottomrule
  \end{tabular}
\end{table}

\begin{table}[h]
  \caption{The 49-syllable performance on Du-IN dataset from subjects (01-04).}
  \label{table:result-duin-syllable-subjs-1}
  \centering
  \tabcolsep=0.01\linewidth
  \begin{tabular}{l|c|cccc}
    \toprule
    \multirow{2}{*}{\textbf{Methods}} & \multirow{2}{*}{\textbf{Chan. Select.}} & \multicolumn{4}{c}{\textbf{Accuracy (\%) $\pm$ std (\%)}} \\
    & & subj-01 & subj-02 & subj-03 & subj-04 \\
    \midrule
    EEG-CFMR \cite{song2022eeg}    & MC & 74.38$\pm$0.91 & 76.74$\pm$1.53 & 42.35$\pm$1.84 & 67.50$\pm$1.88 \\
    Du-IN \cite{zheng2025discrete} & MC & 84.96$\pm$1.07 & 83.34$\pm$1.50 & 55.09$\pm$1.64 & 75.92$\pm$1.56 \\
    H2DiLR \cite{wu2024towards}    & MC & 45.77$\pm$1.80 & 53.68$\pm$1.19 & 34.16$\pm$0.90 & 41.40$\pm$0.96 \\
    \midrule
    BrainStra.-Fine                & MC                & \textbf{86.86$\pm$0.70} & \underline{84.04$\pm$1.11} & \underline{60.29$\pm$0.82} & \textbf{82.32$\pm$2.01} \\
    BrainStra.-Fine                & BrainStra.-Coarse & \underline{84.11$\pm$0.60} & \textbf{84.64$\pm$0.82} & \textbf{65.03$\pm$2.06} & \underline{82.05$\pm$0.72} \\
    \bottomrule
  \end{tabular}
\end{table}

\begin{table}[h]
  \caption{The 49-syllable performance on Du-IN dataset from subjects (05-08).}
  \label{table:result-duin-syllable-subjs-2}
  \centering
  \tabcolsep=0.01\linewidth
  \begin{tabular}{l|c|cccc}
    \toprule
    \multirow{2}{*}{\textbf{Methods}} & \multirow{2}{*}{\textbf{Chan. Select.}} & \multicolumn{4}{c}{\textbf{Accuracy (\%) $\pm$ std (\%)}} \\
    & & subj-05 & subj-06 & subj-07 & subj-08 \\
    \midrule
    EEG-CFMR \cite{song2022eeg}    & MC & 80.43$\pm$0.77 & 50.34$\pm$1.58 & 57.13$\pm$2.27 & 58.97$\pm$1.13 \\
    Du-IN \cite{zheng2025discrete} & MC & 86.45$\pm$1.94 & 60.50$\pm$2.05 & 62.77$\pm$2.57 & 67.67$\pm$0.99 \\
    H2DiLR \cite{wu2024towards}    & MC & 55.72$\pm$1.60 & 40.14$\pm$0.58 & 40.96$\pm$1.36 & 39.69$\pm$1.46 \\
    \midrule
    BrainStra.-Fine                & MC                & \textbf{89.86$\pm$0.86} & \textbf{66.66$\pm$1.45} & \textbf{71.77$\pm$1.31} & \textbf{71.39$\pm$1.14} \\
    BrainStra.-Fine                & BrainStra.-Coarse & \underline{89.57$\pm$0.30} & \underline{65.56$\pm$1.48} & \underline{69.25$\pm$1.40} & \underline{69.76$\pm$1.41} \\
    \bottomrule
  \end{tabular}
\end{table}

\begin{table}[h]
  \caption{The 49-syllable performance on Du-IN dataset from subjects (09-12).}
  \label{table:result-duin-syllable-subjs-3}
  \centering
  \tabcolsep=0.01\linewidth
  \begin{tabular}{l|c|cccc}
    \toprule
    \multirow{2}{*}{\textbf{Methods}} & \multirow{2}{*}{\textbf{Chan. Select.}} & \multicolumn{4}{c}{\textbf{Accuracy (\%) $\pm$ std (\%)}} \\
    & & subj-09 & subj-10 & subj-11 & subj-12 \\
    \midrule
    EEG-CFMR \cite{song2022eeg}    & MC & 72.94$\pm$1.86 & 41.08$\pm$0.78 & 72.52$\pm$1.15 & 59.21$\pm$2.26 \\
    Du-IN \cite{zheng2025discrete} & MC & 78.30$\pm$1.05 & 42.38$\pm$2.81 & 80.72$\pm$1.47 & 69.79$\pm$3.80 \\
    H2DiLR \cite{wu2024towards}    & MC & 46.95$\pm$0.52 & 33.93$\pm$0.88 & 43.90$\pm$0.98 & 43.21$\pm$1.17 \\
    \midrule
    BrainStra.-Fine                & MC                & \underline{83.85$\pm$1.36} & \textbf{52.36$\pm$2.13} & \underline{83.12$\pm$0.80} & \underline{73.82$\pm$3.19} \\
    BrainStra.-Fine                & BrainStra.-Coarse & \textbf{84.49$\pm$1.15} & \underline{50.95$\pm$1.14} & \textbf{84.58$\pm$1.31} & \textbf{74.38$\pm$1.28} \\
    \bottomrule
  \end{tabular}
\end{table}

\clearpage
\section{Subject-Wise Channel Connectivity}
\label{sec:supp-channel-connectivity}
We provide detailed information on the channel connectivity for each subject.
\subsection{Du-IN sEEG dataset}
\begin{figure}[h]
  \centering
  \includegraphics[width=\linewidth]{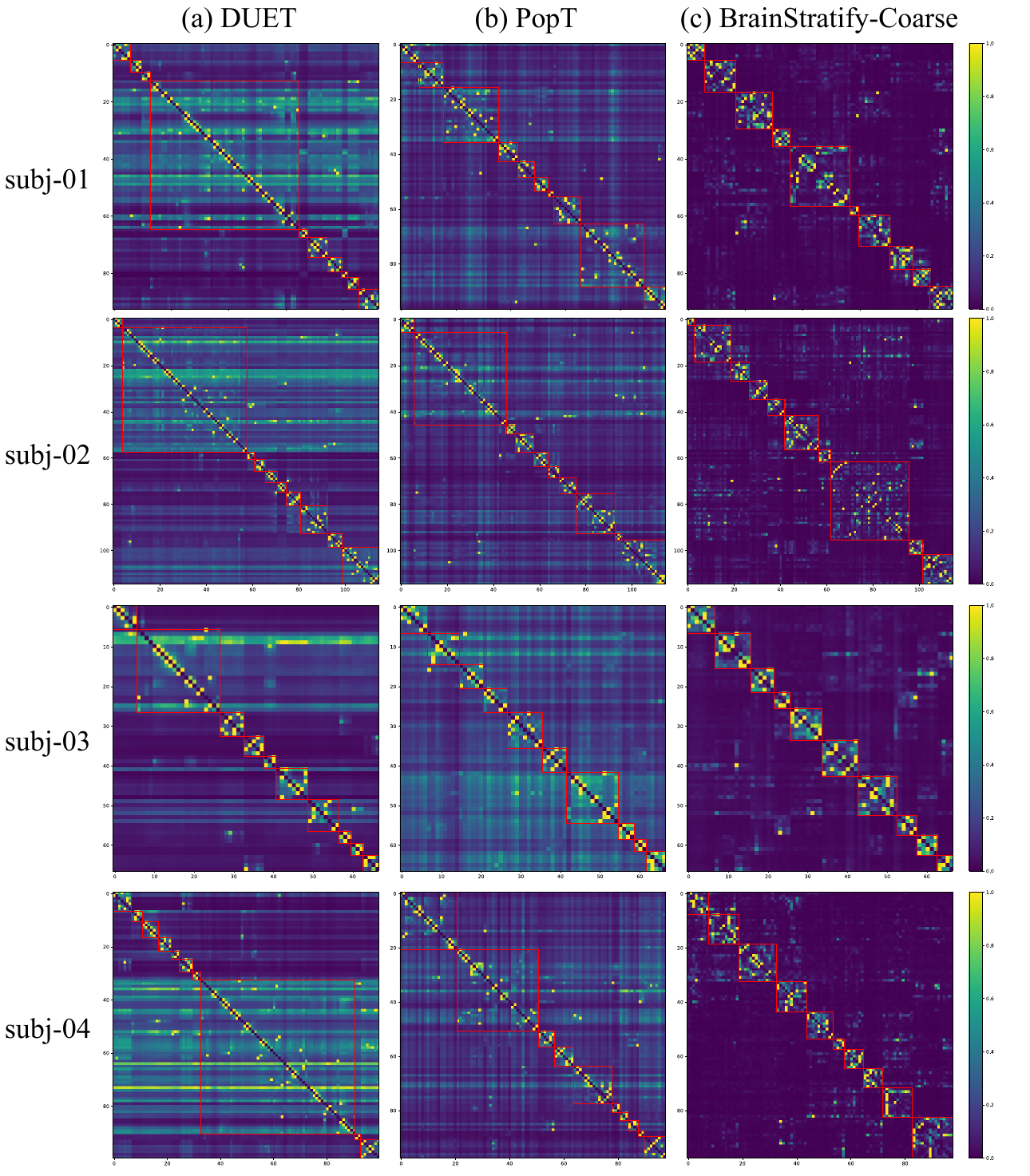}
  \caption{Channel Connectivity from subjects (01-04).}
  \label{fig:channel-connectivity-duin-1}
\end{figure}
\begin{figure}[h]
  \centering
  \includegraphics[width=\linewidth]{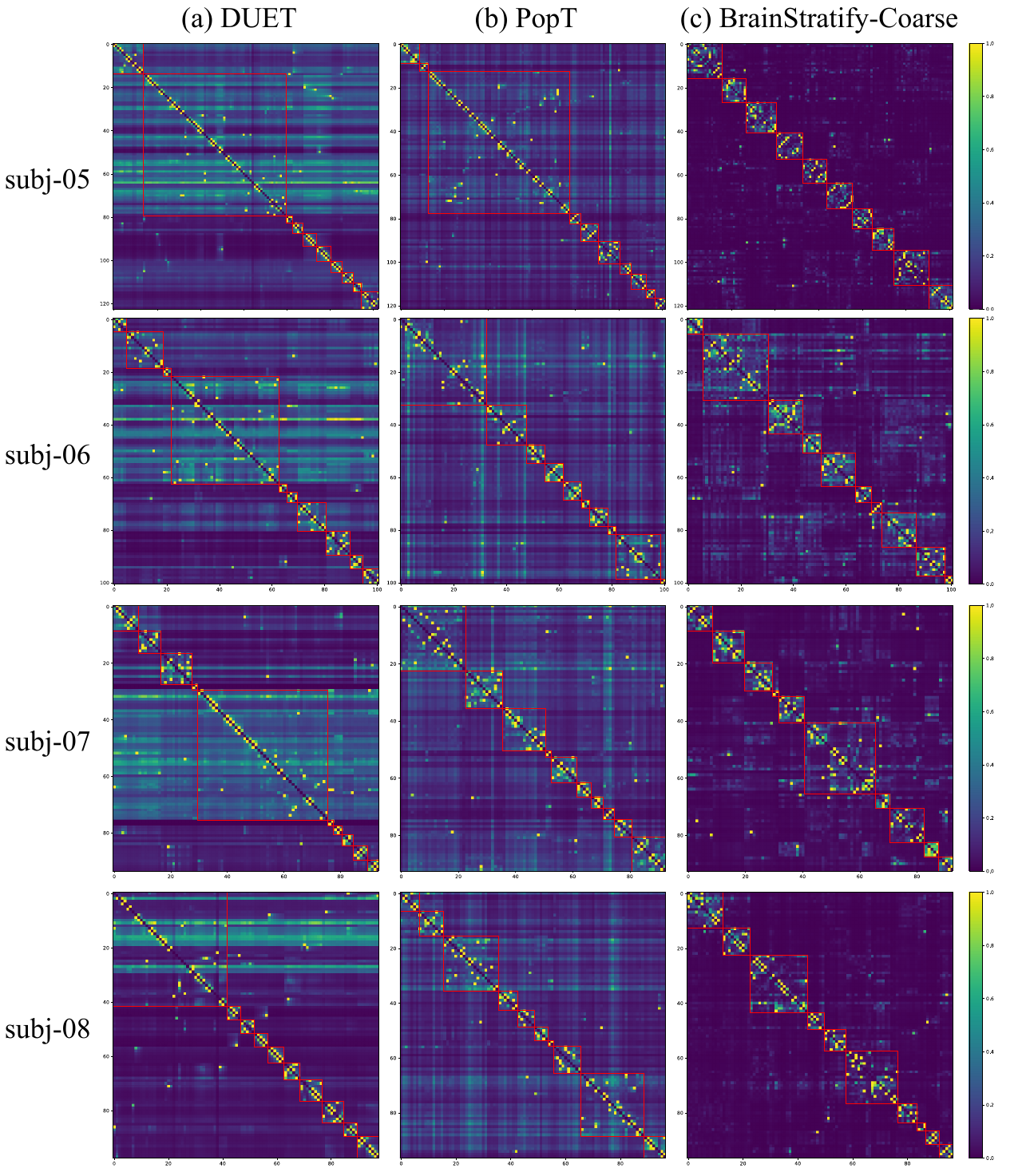}
  \caption{Channel Connectivity from subjects (05-08).}
  \label{fig:channel-connectivity-duin-2}
\end{figure}
\begin{figure}[h]
  \centering
  \includegraphics[width=\linewidth]{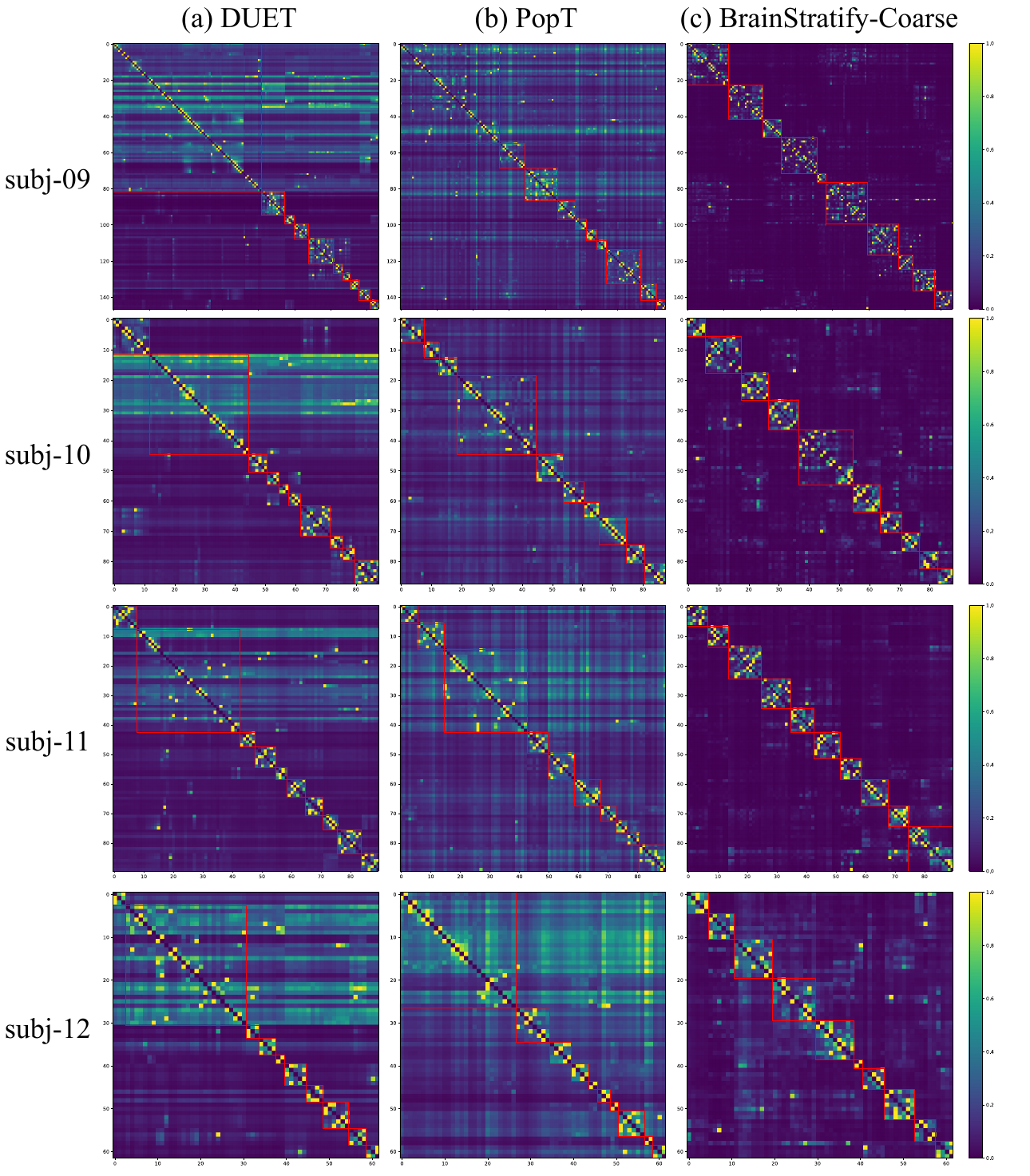}
  \caption{Channel Connectivity from subjects (09-12).}
  \label{fig:channel-connectivity-duin-3}
\end{figure}
\clearpage
\subsection{Brain Treebank sEEG dataset}
\begin{figure}[h]
  \centering
  \includegraphics[width=\linewidth]{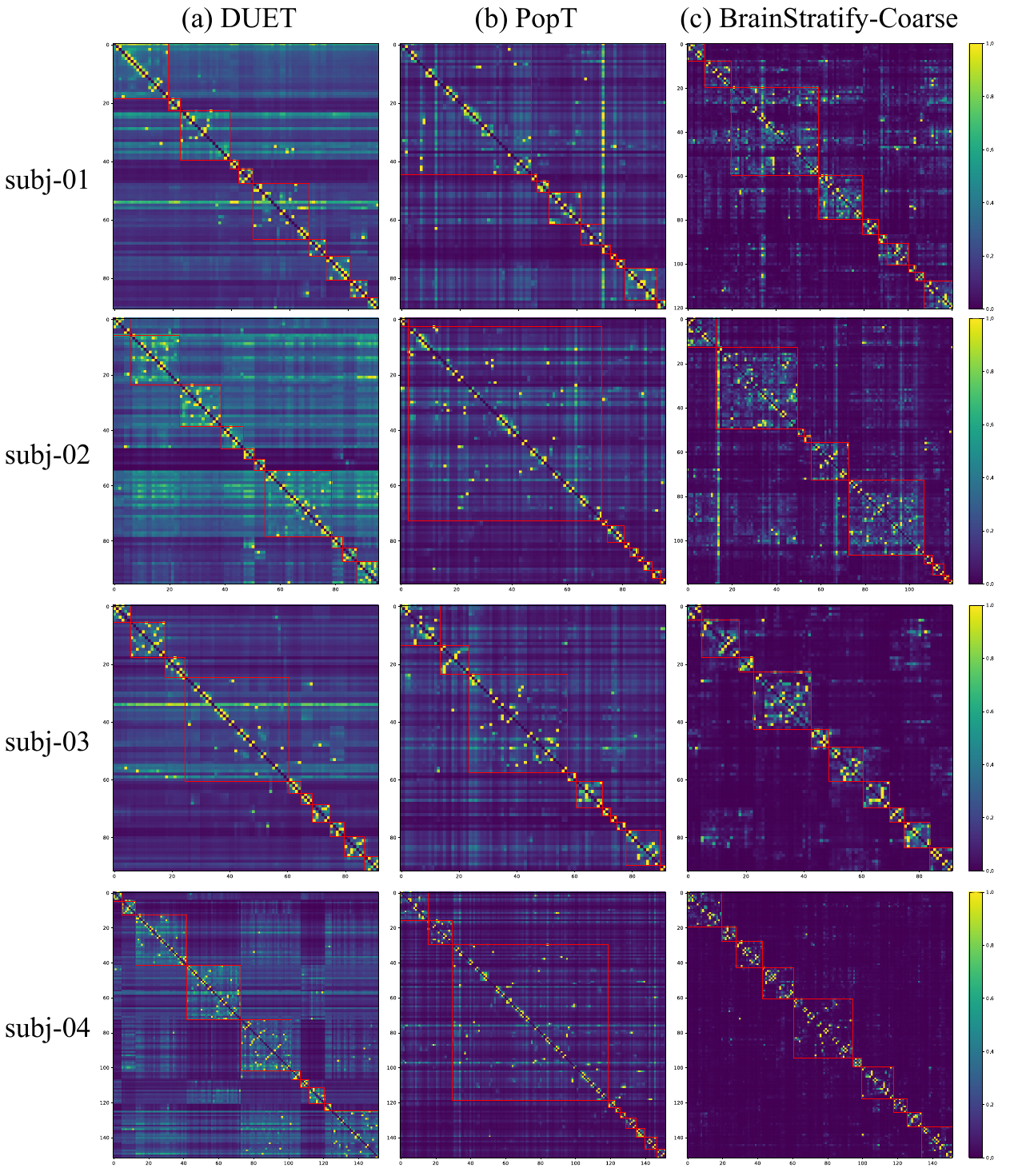}
  \caption{Channel Connectivity from subjects (01-04).}
  \label{fig:channel-connectivity-braintree-1}
\end{figure}
\begin{figure}[h]
  \centering
  \includegraphics[width=\linewidth]{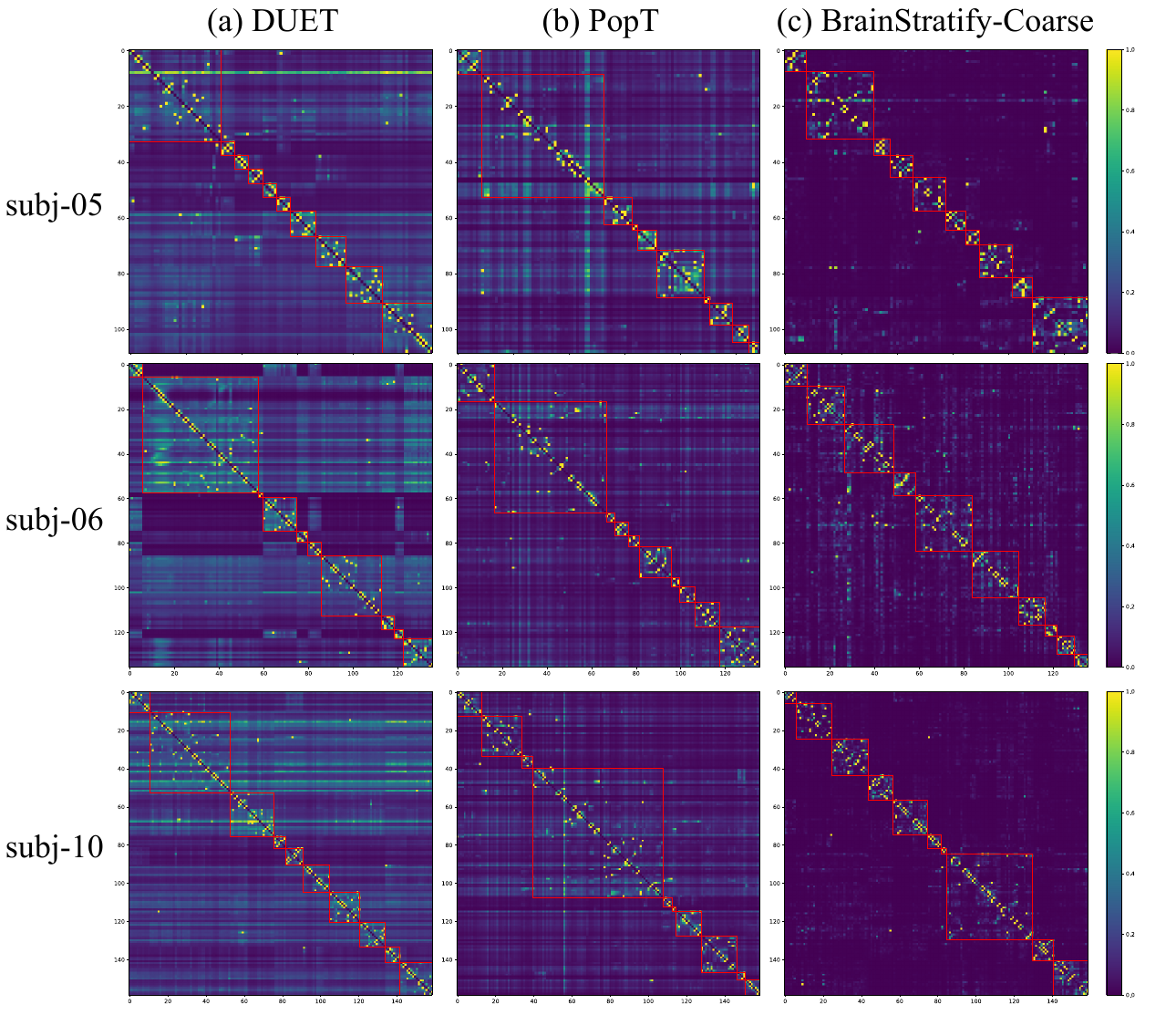}
  \caption{Channel Connectivity from subjects (05,06,10).}
  \label{fig:channel-connectivity-braintree-2}
\end{figure}

\end{document}